\documentclass[a4paper,fleqn,usenatbib]{mnras}
\usepackage{graphicx}
\usepackage{amsmath}
\usepackage{amssymb}
\usepackage{bm}

\usepackage[T1]{fontenc}
\usepackage{ae,aecompl}

\usepackage{txfonts}

\title[Compatibility of the LQGs]{Compatibility of the Large Quasar Groups with the Concordance Cosmological Model}

\author[G. Marinello et al.]{Gabriel E. Marinello,$^1$\thanks{E-mail: gmarinello@uclan.ac.uk, gmarinellob@gmail.com}
 Roger G. Clowes,$^1$ 
 Luis E. Campusano,$^2$
 \newauthor
 Gerard M. Williger,$^{1,3}$ 
 Ilona K. S\"{o}chting$^4$
and Matthew J. Graham$^5$\\
$^1$Jeremiah Horrocks Institute, University of Central Lancashire, Preston PR1 2HE\\
$^2$Observatorio Astron\'{o}mico Cerro Cal\'{a}n, Departamento de Astronom\'{i}a, Universidad de Chile, Casilla 36-D, Santiago, Chile\\
$^3$Department of Physics \& Astronomy, University Louisville, Louisville, KY 40292, USA\\
$^4$Astrophysics, Denys Wilkinson Building, Keble Road, University of Oxford, Oxford OX1 3RH\\
$^5$California Institute of Technology, 1200 East California Boulevard, Pasadena, CA 91125, USA}

\date{\today} 
\pubyear{2016}

\begin{document}
\label{firstpage}
\pagerange{\pageref{firstpage}--\pageref{lastpage}}
\maketitle

\begin{abstract}
We study the compatibility of large quasar groups with the concordance cosmological model. Large Quasar Groups are very large spatial associations of quasars in the cosmic web, with sizes of $50-250h^{-1}$ Mpc. In particular, the largest large quasar group known, named Huge-LQG, has a longest axis of $\sim 860 h^{-1}$ Mpc, larger than the scale of homogeneity ($\sim 260$ Mpc), which has been noted as a possible violation of the cosmological principle. Using mock catalogues constructed from the Horizon Run 2 cosmological simulation, we found that large quasar groups size, quasar member number and mean overdensity distributions in the mocks agree with observations. The Huge-LQG is found to be a rare group with a probability of $0.3$ per cent of finding a group as large or larger than the observed, but an extreme value analysis shows that it is an expected maximum in the sample volume with a probability of 19 per cent of observing a largest quasar group as large or larger than Huge-LQG. The Huge-LQG is expected to be the largest structure in a volume at least $5.3 \pm 1$ times larger than the one currently studied.
\end{abstract}


\section{Introduction}

The spatial distribution of galaxies is not homogeneous, but instead galaxies form a complex hierarchy of structures, forming clusters and groups, filaments and walls of galaxies that surround enormous underdense volumes called voids. Together, they are known as the large-scale structure of the Universe (LSS, hereafter) or the Cosmic Web. The LSS is one of the main subjects of study of observational cosmology as it allows us to probe the underlying distribution of matter of the Universe \citep{Peebles1980}, and it provides fundamental constraints for the concordance cosmological model \citep{Spergel2003}. In particular, the LSS can provide important evidence about the existence of non-gaussianities in the initial conditions or violations of hypothesized homogeneity and isotropy of the Universe, the so-called cosmological principle. However, the study of these ideas requires very large volume surveys, due to the large scales studied. But, redshift surveys using normal galaxies are currently limited to redshifts too low for a good sampling of the very large scales needed to test them. Instead, other kind of objects have been used to sample medium to high redshifts. Quasars, the brightest class of Active Galactic Nuclei (AGN) \citep[see][]{Antonucci1993}, with their high luminosities have allowed the construction of redshift surveys at medium and high redshift, e.g. the 2dF QSO Redshift Survey \citep{Colless2001,colless2003} and the Sloan Digital Sky Survey Data Quasar Catalogue \citep[see][]{Schneider2010}. The larger volumes and redshift ranges available in quasar redshift surveys have produced new opportunities to test deviations from the standard cosmology \citep[e.g.][]{Sawangwit2012} or the cosmological principle.

One of the findings in the quasar large-scale structure is the presence of very large groups or associations of quasars at large scales (hundreds of Mpc), that we call Large Quasar Groups (LQGs). LQGs are very large in comparison with low redshift structures, with dimensions in the range of $50-250$ Mpc, and they have been detected in the redshift range $z \sim 0.4-2.0$. The first detection was done by \citet{Webster1982}, but this first detection was done for a small sample and the redshifts have larger errors in comparison with modern surveys. It was not until the construction of sizeable quasar redshift surveys that it was possible to improve the confidence in the detection, increasing the number of successful detections in the following years. The Clowes-Campusano LQG \citep*{Clowes1991} was the largest LQG known in the literature with a size of $\sim 250$ Mpc (now the largest LQG is the Huge-LQG). The Clowes-Campusano LQG is also the most studied of these structures \citep{Haines2004,Haberzettl2009,Clowes2012,Clowes2013b,Einasto2014}. For further references about LQGs see \citet{Clowes2012} and references therein. LQGs provide a sample of large-scale overdensities at medium redshifts for which the galaxy LSS is not well mapped. Therefore, they could provide a way to study the growth of the large-scale structure, its connection to the galaxy formation, and possibly additional constraints on the concordance cosmological model.

Using the SDSS DR7QSO redshift survey, \citet{Clowes2012,Clowes2013,Clowes2013b} performed a search for LQGs at medium redshifts, and produced a catalogue that represents an order-of-magnitude increase in the number of known LQGs over the pre-SDSS total. \citet{Clowes2012} corroborate the previous detection of the Clowes-Campusano LQG and found a similar detection in the neighbourhood. Therefore this kind of LQG is more common than previously expected. \citet{Clowes2013} shows the detection of the largest LQG known with an extension of $868 h^{-1}$ Mpc, which is known as the Huge-LQG. This structure has received additional corroboration using MgII absorbers \citep{Clowes2013}. \citet{Hutsemekers2014} found that the quasar polarization is partially correlated with the direction of its main branches, which might be an indication of the association of the LQG sub-structures with walls in the LSS \citep[see][]{Hahn2007}. Other groups have recovered similar LQGs using slightly different definitions \citep[e.g.][]{Einasto2014}.

The very large size of the Huge-LQG and other LQGs could represent a challenge for the accepted values of `the scale of homogeneity', the minimum scale at which the Universe looks statistically homogeneous, a fundamental assumption of the standard cosmological model. Estimates for the scale of homogeneity are as small as 60-70$h^{-1}$Mpc \citep{Hogg2005,Yadav2005,Sarkar2009} or as large as 260$h^{-1}$Mpc \citep{Yadav2010}. Similar claims of extreme structures in the LSS have been made before, as the case of the Sloan Great Wall \citep{Gott2005} in the SDSS redshift survey, and two large ``hotspots'' that correspond to two superclusters in the 2dFGRS \citep{Baugh2004,Croton2004}. These were found to be consistent with the LSS in the concordance cosmology by \citet{Yaryura2011} (study of the two large superclusters in the 2dFGRS volume), \citet*{ShethDiaferio2011} (study of the compatibility of Shapley supercluster and the Sloan Great Wall) and \citet{Park2012} (analysis of the Sloan Great Wall compatibility using cosmological simulations).

There have been performed various analyses about the likelihood of the Huge-LQG \citep{Pilipenko2013,Nadathur2013,Park2015}. However, all these analyses are based on random catalogues, samples of randomly distributed points from a uniform distribution in the volume. They are therefore just testing the statistical significance of a LQG against a null hypothesis of complete spatial randomness, instead of compatibility with the expected LSS. \citet{Nadathur2013} performed an analysis of the homogeneity of the quasar sample, which he found to be consistent with homogeneity, and an analysis of the likelihood of the Huge-LQG in random catalogues, using the probability that the largest random group has more members than Huge-LQG. Using the same linking length chosen by us, he finds that this p-value probability is $0.085$. However, this is not a significance test for the group but an extreme value analysis as it deals with the largest object in the sample. A true significance test has to compare it against a random group population. \citet{Park2015} used a similar analysis using length and richness. No extreme value or outlier analysis is given, in this case. In their conclusion they correctly say the existence of the Huge-LQG does not directly imply a challenge for the concordance cosmological model and further analyses using cosmological simulations are needed. Hence, there has not been a proper comparison with the LSS predicted by the concordance model. This comparison is necessary as random catalogues cannot give any insight into the relation between the LSS and the LQGs.

In this paper, we performed an analysis of the LQG compatibility with the concordance cosmological model by comparing the observed LQGs with those from mock LQG catalogues obtained from the Horizon Run 2 cosmological simulation \citep{Kim2009}. These LQG mock catalogues were constructed from a set of mock quasar catalogues using the same group finder and significance test employed for the observational sample. We also perform an extreme value analysis of the probability of observing the Huge-LQG in these mock catalogues.
			
The paper is organized as follows. In section \ref{sec:obslqg}, we describe the observational sample of LQGs used in the comparison and we discuss its main properties. In section \ref{sec:mockqso} we describe the construction of the intermediate mock quasar catalogues from the Horizon Run 2 cosmological simulation. In section \ref{sec:mocklqg} we show the resulting mock LQG catalogues and their comparison against observations. In section \ref{sec:eva} we show the extreme value analysis of the largest LQG in size and quasar number. Finally, in the last section we present the summary and discussion of our results.

We adopt a fiducial cosmological model close to the best values of the concordance cosmological model, with $\Omega_T = 1$, $\Omega_\mathrm{M,0} = 0.27$, $\Omega_{\Lambda,0} = 0.73$ and $H_0 = 70\ \rmn{km}\ \rmn{s}^{-1}\ \rmn{Mpc}^{-1}$. All distances shown are comoving distances.

\section{Large quasar group sample}
\label{sec:obslqg}

\begin{figure*}
\centering
\includegraphics[width=0.45\textwidth]{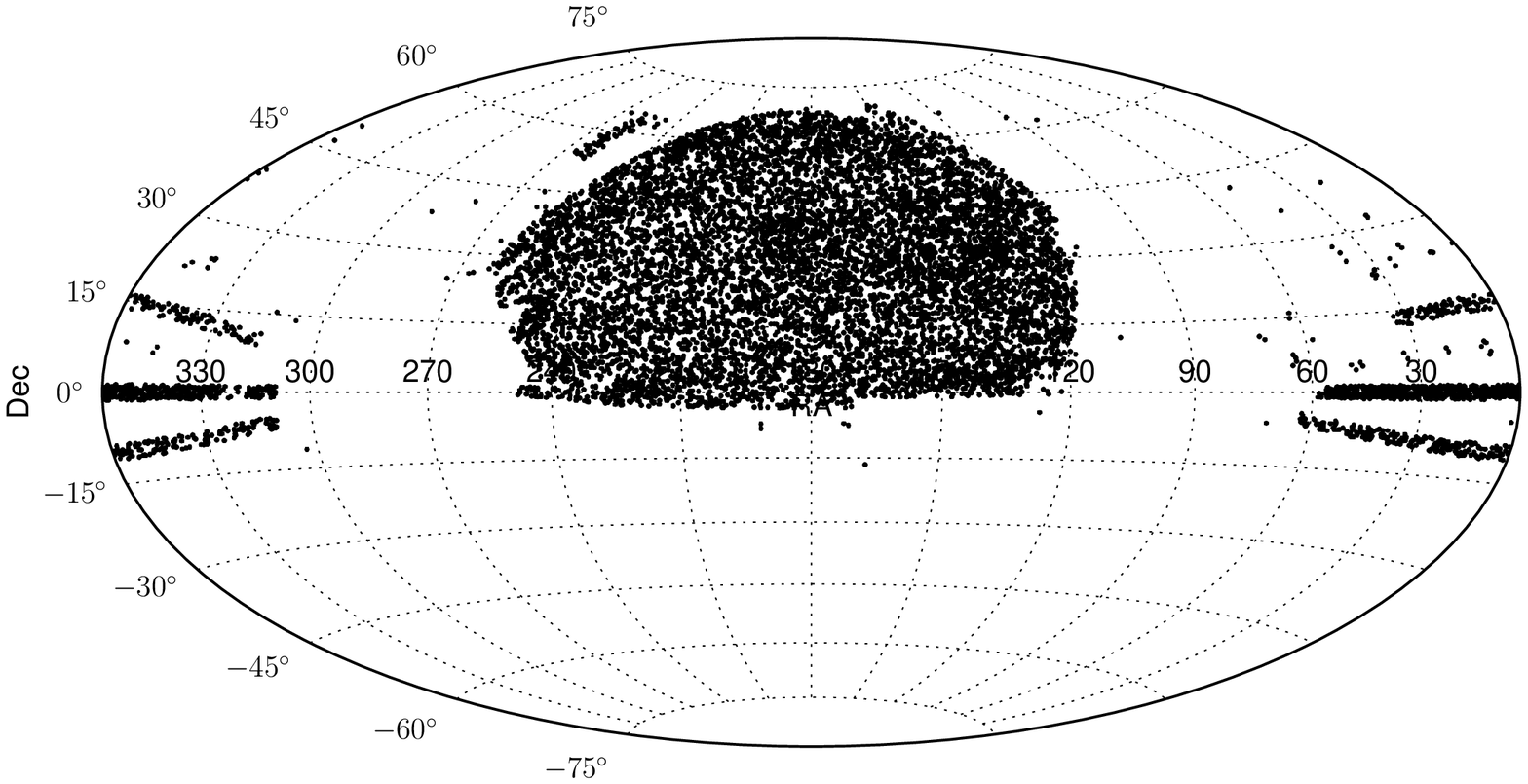}
\includegraphics[width=0.45\textwidth]{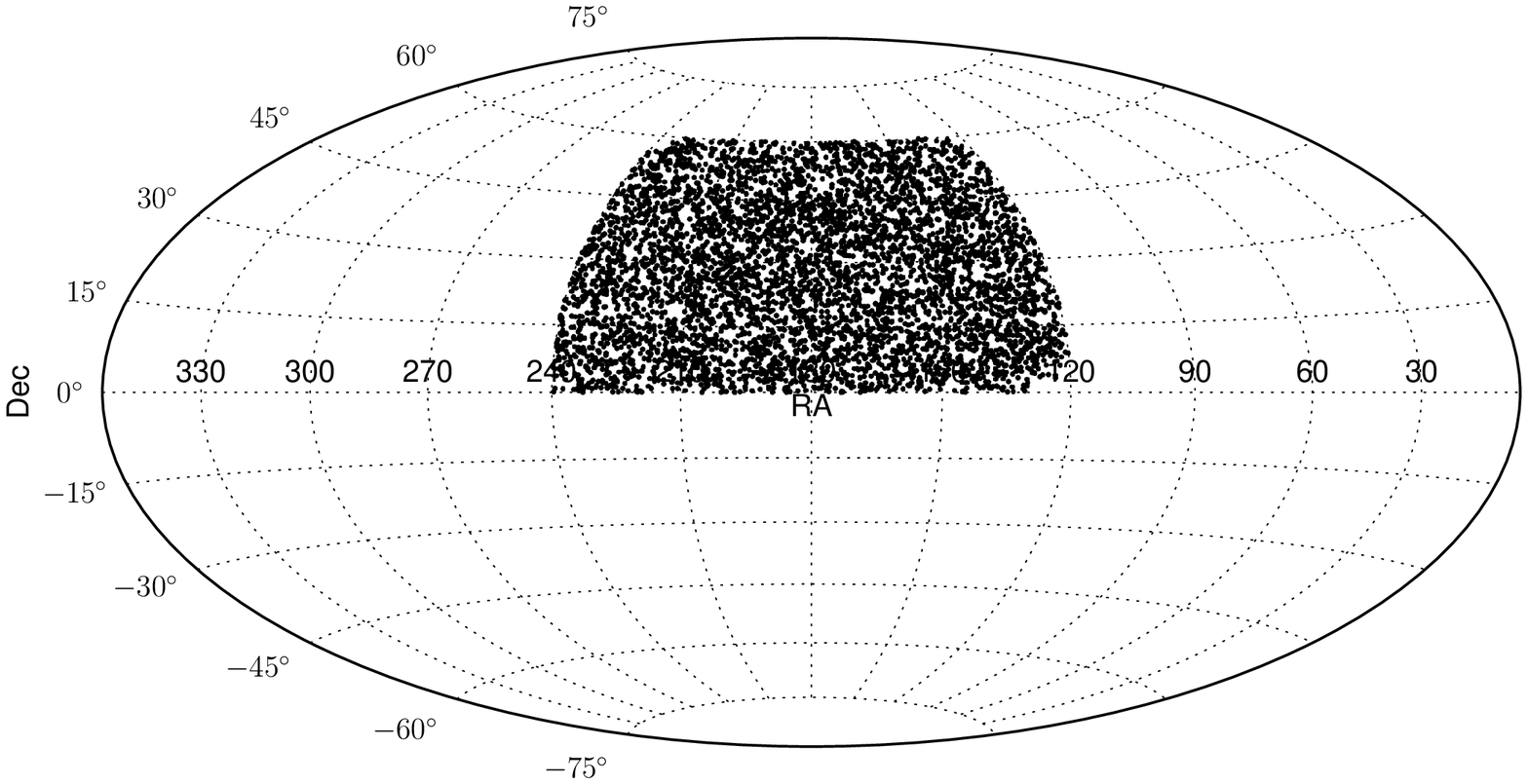}
\caption[Hammer-Aitoff sky projection of the full SDSS Quasar Redshift Survey DR7]{Hammer-Aitoff sky projection of the full SDSS Quasar Redshift Survey DR7 (SDSS-QSO DR7, left panel) and our sample (right panel). For display purposes, only a random selection of the total sample is shown. For the full catalogue a tenth of the quasars is shown, and half of the quasars in the the quasar sample. The largest contiguous region corresponds to the SDSS Legacy survey. The angular limits restricted the quasar sample to a region that contains most of the contiguous region of the SDSS-QSO DR7, avoiding the regions with very low completeness in the outskirts of the SDSS survey.}
\label{fig:sampleprojection}
\end{figure*}

The Large Quasar Group catalogue in \citet{Clowes2012,Clowes2013} was constructed using the SDSS Quasar Redshift Survey Data Release 7 \citep[][SDSS-QSO DR7]{Schneider2010}. This catalogue is part of the Sloan Digital Sky Survey \citep{York2000} and was the largest quasar redshift survey at the time, being the conclusion of the SDSS-I and SDSS-II quasar surveys. The catalogue contains 105,783 spectroscopically confirmed quasars with luminosities brighter than $M_i = -22.0$ (in their fiducial cosmology) and fainter than $i \approx 15.0$. Most of these quasars have highly reliable redshifts with errors of the order of a few percent. The catalogue covers an area of $\sim 9380\ \mathrm{deg}^2$. The quasar redshifts range from $0.06$ to $5.46$, with a median value of $1.49$. The spatial distribution of the survey is shown in Fig. \ref{fig:sampleprojection}. This catalogue does not constitute a statistical sample, i.e.. a sample obtained through homogeneous selection, as noted by \citet{Schneider2010}. \citet{Richards2006} describe how to construct a statistical sample from the DR3QSO catalogue. However, their criteria were chosen to estimate the luminosity function of the sample not for reconstruction of structures, so it can be modified if necessary.  The quasar sample chosen by \citet{Clowes2012,Clowes2013} is defined by those quasars with apparent magnitude $i \leq 19.1$ for the entire angular coverage of the catalogue and redshift in the range $1.0 \leq z \leq 1.8$. The magnitude limit gave an approximately spatial uniform selection for redshifts $z \leq 2$ \citep{VandenBerk2005,Richards2006}. This sample is called A9380, from its solid angle in squared degrees. Additionally, a control area, called A3725,  was defined in RA $123 \fdg 0 - 237 \fdg 0$ and Dec. $15\fdg0 - 56\fdg0$, which encloses $\sim 3725$ deg$^2$. This region is used in the statistical analysis of the LQG candidates. Its limits were chosen so the region does not include the Clowes-Campusano LQG, the largest LQG known before \citet{Clowes2013}, a possible outlier in the quasar distribution that could possible bias the mean quasar density estimation. The control area also does not include the Huge-LQG, therefore is still a valid control area in this regard.

However, this sample is not completely adequate for comparing the LQGs against cosmological simulations, as this was not the main goal in the previous papers. The redshift distribution in this redshift range is flat and therefore the quasar number density is decaying with redshift. This effect is mainly due to the magnitude limit of the sample, but also it is caused by the evolution of the luminosity function, which can be explained by either luminosity evolution or density evolution \citep[see][]{Richards2006}. This variation affects the group finder as the algorithm used, the Friends-of-Friends algorithm, depends on the local density. As a result, the groups at higher redshift are fewer and smaller than lower redshift groups for a fixed linking length or threshold density. Additionally, the clustering properties of quasars change with redshift and the bias parameter increases from $\sim 2.0$ at redshift $1$ to $\sim 3.0$ at redshift $2$ \citep{Porciani2004,Ross2009}. This effect can be explained if quasars are hosted in massive dark matter haloes, of $\sim10^{13} \mathrm{M}_\odot$, and there is not time evolution in the characteristic mass of these haloes (see subsection \ref{subsec:hodmodel}). In this case, the evolution in the quasar clustering introduces a systematic effect in the detection of LQGs as the variance of the quasar sample changes with radial distance to the observer. This effect makes harder the construction of mock catalogues as we need many snapshots of a simulation at different redshifts. To perform a comparison with simulations we rerun the group finder in a subset of the original volume used in \citet{Clowes2012,Clowes2013} so the LQGs are more uniformly detected. We use the redshift range $1.2 \leq z \leq 1.6$, which roughly corresponds to an interval of one billion years in cosmic time. In this range the clustering evolution is weak enough that the change in the estimated bias is within its error \citep[][]{Porciani2004, Ross2009}. We restrict the sample to RA $120\fdg0 - 240\fdg0$ and Dec. $0\fdg0 - 60\fdg0$, avoiding regions with very low completeness in the SDSS catalogue that could affect the group finder performance. The angular distribution of this quasar sample is shown in Fig. \ref{fig:sampleprojection}.

We employed the same methodology described in \citet{Clowes2012} to construct the LQG catalogue. We summarize the procedure here. A detailed description can be found in \citet{Clowes2012}. We use the Friend-of-Friends method \citep*[FOF][]{PressDavis1982} as the group finder.  Also called single-linkage hierarchical clustering in the statistical literature, this is a well-known cluster finder. It is related to the Minimal Spanning Tree (MST) \citep{kruskal,prim}, a construct of graph theory, as FOF clusters can be obtained from the division or `pruning' of a MST at a certain length. The FOF algorithm links together in the same group all objects with mutual separation less than a certain distance, usually called the \textit{linking length}. Therefore, the linking length must be chosen carefully in order to avoid the risk of \textit{percolation}, the merging of most objects in a single group spanning the volume. A possible specification of the linking length is the mean nearest-neighbour separation of the sample, $\sim 74$ Mpc, which is a good compromise value in general. But this value is estimated in redshift space. By including the effect of redshift errors and peculiar velocities in the estimation of comoving distances, and testing the detection performance a comoving linking length of 100 Mpc in our fiducial cosmology  was selected. We do not consider candidate LQGs with fewer than 10 members because groups with fewer than 10 members are more likely to be affected by noise and their geometrical properties are very uncertain.

Once we obtained the candidate groups we performed a hypothesis test against the null hypothesis of groups come from a point set that is randomly distributed, i.e. they correspond to random groups. The statistic used is the volume of the point distribution estimated using a procedure we called the convex hull of member spheres \citep[][CHMS, hereafter]{Clowes2012}. This method assigns a sphere with radius half of the mean linkage (MST edge length) to each member position and then computes the convex hull of the sphere set. This method is conservative and avoids underestimations due to the use of single points in the estimation of the convex hull. As random groups have larger expected volumes than real groups, we define one-sided test with a specified region of rejection for group volumes larger than a critical volume. We chose this region to represent a significance level of 2.5 per cent. For a one-sided test and assuming approximate normality this level correspond to $2.8$ standard deviations from the mean. The critical volume depends on the number of members, therefore for each quasar-member number the region of rejection is estimated using Monte Carlo simulation of random groups from uniformly distributed points with the same number density.

\begin{figure}
\label{fig:lqgradec}
\centering
\includegraphics[width=0.5\textwidth]{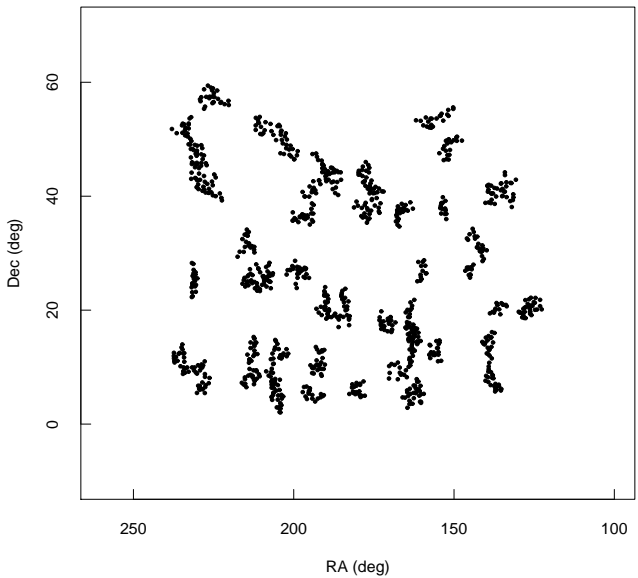}
\caption[RA-Dec plot of the LQG member quasars.]{RA-Dec plot of the LQG member quasars. LQGs are found to surround low-density regions, delimiting cosmic voids. The geometry of the LQGs is web-like and elongated.}
\end{figure}

The final LQG sample then consists of 59 significant groups (from 189 candidates). The distribution of these LQGs is shown in Fig. \ref{fig:lqgradec}.

\section{Mock quasar catalogues}
\label{sec:mockqso}

To predict the properties of the LQG population in the standard cosmology it is necessary to be able to perform the same procedure as used in the observational catalogue, so that the resulting mock catalogues have the same biases and systematics errors present in observations. Therefore, it is necessary to construct a set of mock LQG catalogues from a cosmological simulation. As LQGs are constructed from the quasar distribution, it is also necessary to construct intermediate mock quasar catalogues that reproduce the same biases and selection effects found in the SDSS-DR7 quasar catalogue. In this section we describe the procedures followed to construct these intermediate mock quasar catalogues from the chosen cosmological simulation and their testing against observations.

\subsection{Cosmological simulation}

As LQGs are very large structures we require that the cosmological simulation selected is accurate at large scales. Therefore, the volume of the simulation needs to be many times larger than the largest structure, because the largest wavelength perturbations detectable are equal to the box size ($L$) of the simulation. Additionally, high resolution to sub-galaxy scales is not necessary because of the current uncertainties in the theory of the AGN formation and evolution makes necessary the use of analytical approximations at galaxy scales anyway. Also, the observational errors in the radial and angular position and the sampling bias from the fibre collisions means that the quasars can only be sampled with an accuracy 1 Mpc at the typical redshift of the sample, which correspond to the size of a galaxy cluster \citep{Weygaert2006}. Therefore, we need a detailed simulation only at the level of dark matter haloes. Haloes are collapsed regions in the matter field and they have a dynamical quasi-equilibrium distribution that is thought to be universal \citep[see ][for a review]{CooraySheth2002}.

We selected the Horizon Run 2 cosmological N-body simulation \citep{Kim2011}. Horizon Run 2 (HR2, hereafter) is one of the largest pure dark matter cosmological simulations available, with $6000^3=216$ billion particles, and a volume of ($7200 h^{-1}$ Mpc)$^{3}$, which is more than 2700 times the volume of the Millennium Run \citep{springel2005}. It resolves galaxy cluster size haloes with mean separations of $1.2 h^{-1}$ Mpc, and the minimum halo mass is $3.75\times 10^{12} h^{-1} \mathrm{M}_{\odot}$. A summary of the main parameters of the HR2 simulation is shown in Table \ref{tab:hr2}. HR2 is a dark-matter-only simulation, and therefore we use their dark matter halo catalogues to produce mock catalogues. We used the Friends-of-Friends (FoF) dark matter halo catalogue of a snapshot of the HR2 at redshift $1.4$, which corresponds to the median redshift of our quasar sample. In principle, a more realistic light-cone simulation should be used, but HR2 does not have the temporal resolution at redshifts higher than one to construct a light-cone mock catalogue. However, the evolution of the dark matter halo clustering in the redshift range $1.2-1.6$ is small, and most of the density, and luminosity, and possible clustering evolution can be ascribed to quasar evolution.


A possible shortcoming of this halo catalogue is that the minimum number of mass particles per halo is low, 30 particles. Therefore, the estimated mass of each halo has an important uncertainty because of Poisson noise. The low number of particles also produces artificial discreteness in the mass distribution that could invalidate the application of analytical formulae that depend on the continuity in halo mass. To avoid this effect, we smooth the halo mass by applying a random scatter following a uniform distribution of half a particle to simulate a continuous distribution of mass levels. Another shortcoming is that FoF haloes tend to include matter particles that are not actually gravitationally bound to the haloes. This affects in particular the high mass end of the mass function of haloes, and as a result the halo mass function is more heavy-tailed than it should be for real virial masses. We correct for this effect by applying the \citet{Warren2006} correction for the number of mass particles, which is based on the comparison of virial and FOF masses in dark matter N-body simulations. The Warren correction is
\begin{equation}
N_{h} = N^\mathrm{FOF} \left( 1 - \left( N^\mathrm{FOF}\right)^{-0.6} \right),
\end{equation}
where $N_h$ is the corrected number of particles and $N^\mathrm{FOF}$ is the number of particles obtained using the FOF method. The corrected mass is $M_h = m_{p} N_h$, where $m_p$ is the particle mass, equal to $1.25 \times 10^{11}h^{-1} \mathrm{M}_{\odot}$ for HR2 and $1.79 \times 10^{11} \mathrm{M}_{\odot}$ in our fiducial cosmology.

\begin{table}
\caption{Main parameters of the Horizon Run 2 simulation.}
\begin{center}
\begin{tabular}{l|l}
\hline
Parameter & HR2 value\\
\hline
Model & WMAP5\\
$\Omega_{\mathrm{M}}$ & $0.26$\\
$\Omega_{b}$ & $0.044$\\
$\Omega_{\Lambda}$ & $0.74$\\
Spectral index & $0.96$\\
$H_0$ [$\mathrm{km}\ \mathrm{s}^{-1}\mathrm{Mpc}$] & $72$\\
$\sigma_8$ & $0.794$\\
Box size [$h^{-1} \mathrm{Mpc}$] & $7200$\\
Number of particles & $6000^3$\\
Starting redshift & 32 \\
Particle mass [$10^{11} h^{-1} \mathrm{M}_{\odot}$]& $1.25$\\
Mean particle separation [$h^{-1} \mathrm{Mpc}$]& $1.2$\\
Minimum halo mass (30 particles)[$10^{11} h^{-1} \mathrm{M}_{\odot}$] & $37.5$\\
\hline
\end{tabular}
\end{center}
\label{tab:hr2}
\end{table}

The volume of the Horizon Run 2 simulation is divided into subvolumes with the same geometry as the sample volume. We obtained 11 independent volumes for the construction of the mock catalogues.

\subsection{Quasar Halo Occupation Distribution model}
\label{subsec:hodmodel}

We construct our mock quasar catalogues using a Halo Occupation Distribution \citep*[HOD hereafter,][]{Berlind2002}. This is a probabilistic model of the galaxy (or a different object like quasars) distribution within a dark matter halo. Together with the halo model this completely defines the large-scale distribution of a particular type of galaxy, such as quasars. The HOD can be calibrated using observations and therefore it is particularly suitable for the construction of large mock catalogues. HOD models have been used extensively in the analysis of the clustering of quasars, and they have been successful in clarifying the relation between the dark matter halo distribution and quasars \citep[e.g.][]{MartiniWeinberg2001,Croom2004, Porciani2004,Ross2009,Shen2007,Richardson2012}. These studies found that quasar clustering is consistent with quasars being hosted by massive dark matter haloes of about $10^{12}-10^{13} \mathrm{M}_{\odot}$, independently of redshift. Haloes of this mass range are about $100-1000$ times more abundant than quasars (depending on the model), which excess is explained by quasars having a relatively short duty cycle. Assuming that all haloes with more than the threshold mass contain a galaxy with a SMBH, then the observed fraction of quasars is just the mean duty cycle. For a duty cycle of $0.1-1$ per cent the mean quasar lifetime should be in the range $10-100$Myr, consistent with theoretical predictions of the e-folding time of the SMBH growth \citep*{MartiniWeinberg2001}. This lifetime is an average for a given time period: the actual quasar activity could consist of a series of luminous periods with added total time equal to the mean lifetime, as is suggested by recent evidence supporting this picture \citep[see][for a review]{AlexanderHickox2012}.

We adopted a modification of the \citet*{Berlind2002} HOD model, a widely used HOD model that has been used successfully in galaxy populations, by multiplying the occupation number (mean number of objects in the halo) by a constant effective quasar duty cycle, similarly to  \citet{Padmanabhan2009}. This is the light-bulb model: quasars are either active or inactive, and they radiate at close to peak luminosity. It is widely used in studies of quasar clustering \citep[e.g.][]{MartiniWeinberg2001,Porciani2004,Padmanabhan2009,Croton2009,ConroyWhite2013}. Even though this model is not enough to reproduce the luminosity function and other properties of AGN in general, it is successful in reproducing the spatial clustering of quasars. A constant duty cycle means that the quasar lifetime is independent of the mass of the host halo. Observational evidence shows that the fraction of galaxies with active AGNs increases with stellar mass \citep{Best2005} and correlates with virial halo mass \citep{Behroozi2013}. But in the case of luminous quasars the stellar masses are larger than $10^{11} \mathrm{M}_{\odot}$, and in this regime the fraction is weakly dependent on the stellar mass. Thus, a constant duty cycle is a good approximation of the behaviour of quasars in high mass haloes \citep[e.g.][]{ConroyWhite2013}. This behaviour is related to the triggering mechanism of quasars, that is most likely to be caused by major mergers (see discussion in the following paragraph).

The \citet*{Berlind2002} HOD model distinguishes between central and satellite galaxies, to reproduce the different kinematics and morphologies observed in cluster galaxies. Evidence suggests that low-to-moderate luminosity AGN are hosted in disc-dominated galaxies \citep{Gabor2009,Cisternas2011,Schawinski2011,Kocevski2012} suggesting a predominantly secular fuelling of the AGN by disc instabilities or minor mergers \citep{Hopkins2005,Menci2014}. In contrast, there is strong evidence suggesting that major mergers, i.e. merger of galaxies with similar masses, are the predominant mode of triggering high luminosity AGN as quasars \citep{Kauffmann2003,Treister2010,Treister2012,Villar-Martin2011,Villar-Martin2012,Menci2014}. Also, there is evidence that quasars at all redshifts are preferentially located in early-type galaxies, or at least galaxies with a dominant spheroidal component \citep{Kauffmann2003,Kocevski2012} with respect to normal galaxies, consistent with the relation between the black hole mass and the spheroid mass \citep{Magorrian1998,FerrareseMerritt2000,Tremaine2002,MarconiHunt2003}. There is some evidence that AGN and quasars are preferentially located in satellite galaxies at $z<1$ \citep[][]{sochting02,AlexanderHickox2012}. However, there is not enough evidence to have a clear picture of the location of quasars in dark matter haloes at high redshift. Major mergers are more prevalent in satellite galaxies, but they are also present in central galaxies at high redshift and the formation of central elliptical galaxies requires major mergers of massive galaxies. Therefore, we assume for this work that quasars can be triggered either in central or satellite galaxies, with no strong dependence on the mass of the dark matter halo, as is expected if major mergers are the predominant mode of triggering and fuelling of quasars. This assumption is found in most of the HOD models applied to quasars \citep{Porciani2004,Richardson2012,Padmanabhan2009}.

The main parameter of HOD models is the mean occupation number $\langle N|M \rangle$, the mean number of objects in a halo of mass $M$. After the mean occupation number is defined the actual number of objects is defined by a probability distribution with a parameter that is the mean occupation number. Following \citet*{Berlind2002}, the central quasar mean occupation number is modelled as a step function (this produces the light bulb behaviour),
\begin{equation}	
\langle N|M \rangle_{\rm{cen}} =
\begin{cases}
f_{q} &,M \geq M_{\mathrm{min}}\\
0 &,M < M_{\mathrm{min}}.
\end{cases}
\end{equation}
where $M_{\mathrm{min}}$ is the minimum halo mass for which a dark matter halo can host a quasar, and $f_q$ is the effective duty cycle, i.e. the fraction of haloes that host \textit{active} quasars combined with the effect of the incomplete sampling of quasars. The satellite quasar mean occupation number is a broken power law \citep*[the same as satellite galaxies, see][]{Berlind2002},
\begin{equation}\label{eq:hodmodel}
\langle N|M \rangle_{\rm{sat}}= 
\begin{cases}
f_{q} \left(\frac{M}{M_s} \right)^{\alpha} &,M \geq M_{\mathrm{min}} \\
0 &,M < M_{\mathrm{min}}.
\end{cases}
\end{equation}

The actual number of quasars is obtained from Monte Carlo simulation from the central and satellite probability distributions that defines the HOD. Following \citet*{Berlind2002}, we use a Bernoulli distribution (nearest neighbour distribution) with a success probability equal to $\langle N|M \rangle_{\rm{cen}}$ for the central quasars and a Poisson distribution with rate equal to $\langle N|M \rangle_{\rm{sat}}$ for satellites.

There are four parameters in the model: the minimum halo mass $M_{\mathrm{min}}$, the power law scale $M_s$, the power law $\alpha$ and the fraction of quasars in haloes $f_{q}$. However, we lack enough constraints to accurately fit these parameters and also we want to avoid overfitting of the model. We adopt the scaling relation for the power law index and the power law scale from \citet{Kravtsov2004} to fix $\alpha$ and $M_s$ in our HOD model. They estimated the HOD of sub-haloes: haloes merged into a larger halo as part of the hierarchical formation of structure in dark matter N-body simulations. As it is expected that galaxies are formed in sub-haloes, and their number and kinematics should approximately follow galactic ones \citet{Kravtsov2004} used the \citet*{Berlind2002} HOD to model the distribution of sub-haloes in dark matter haloes. They found that the power law index is approximately $\alpha = 1$ and that the power law scale, $M_{s}$, follows the scaling relation $M_{s} \approx 20 M_{\text{min}}$ at $z=1$.

The two remaining free parameters $M_{\mathrm{min}}$ and $f_{q}$ are fitted using the observed  mean number density and two-point correlation function. The best fit parameters are $M_{\mathrm{min}} = 6.16 \times 10 ^{12} \mathrm{M}_{\odot}$ which is similar to the minimum halo mass found in previous studies \citep{Porciani2004,Richardson2012}, and $f_{q} = 0.002$. Consequently, $M_{s}=20M_{\mathrm{min}}=1.2 \times 10^{14}\mathrm{M}_{\odot}$. The minimum halo mass sets the effective bias of the mock catalogues and therefore it sets their two-point correlation function, whereas the effective duty cycle sets the observed quasar mean number density. The estimated quasar lifetime implied by $f_{q}$ is $t_q=f_q t_H(z=1.4)= 12.7\ \mathrm{Myr}$ for our fiducial cosmology ($\Omega_m = 0.3$, $\Omega_\Lambda = 0.7$, $h=0.7$), where $t_H$ is the Hubble time at mean redshift of the comparison sample ($1.4$). This is consistent with other estimations of the quasar lifetime in the literature \citep[e.g.][]{MartiniWeinberg2001,Richardson2012,ConroyWhite2013}.

The spatial location of a quasar within the halo is not relevant for the mock LQG catalogues, given the scales involved. However, it can be important in the reproduction of selection effects as redshift distortions. We assign a position and velocity to each quasar depending on whether it is a central quasar or a satellite quasar. Central quasars are hosted by the central galaxy and we assume that this is the most massive galaxy in the halo, and therefore it should be located close to the centre of mass of the halo. Consequently, we assign the coordinates and peculiar velocity of the halo itself to the central quasar. Satellite quasars are located in satellite galaxies that are distributed according to the sub-halo spatial distribution. A good model in this case is to assign to each satellite a random position, following the observed radial profile. The radial profile of satellite quasars is found to be well described by a power law \citep{Degraf2011,Chatterjee2012,Chatterjee2013}. \citet{Chatterjee2013} find the the radial profile is well fitted by
\begin{equation}
n(R) = 10^{-0.67} \left( \frac{R}{R_{200}} \right)^{-2.3},
\end{equation}
where $R_{200}$ is the radius within which the enclosed mean density is 200 times the critical density. We then assign a peculiar velocity equal to the centre of mass peculiar velocity plus a random peculiar velocity following the Maxwell distribution. 

\subsection{Selection effects and observational errors}

The SDSS-QSO quasar sample contains many different sources of observational error and biases that need to be reproduced in the mock catalogues, so they can be directly compared against observations. In the following, we detail the main selection effects and the reproduction of these in the mock quasar catalogues.

\subsubsection{Radial selection function: luminosity assignment using abundance matching }

The SDSS-QSO quasar sample is a magnitude limited sample and thus it has a radial selection function. To reproduce the observed radial selection function we assign absolute magnitudes to each quasar using the scheme called Halo Abundance Matching \citep{Kravtsov2004,Tasitsiomi2004,Vale2004,Conroy2006,ConroyWechsler2009,Guo2010,Behroozi2010}. The Halo Abundance Matching (HAM) is a non-parametric estimator of the relation between halo mass and the object luminosity. This method ensures that the resulting mock catalogue has the correct luminosity function and avoids the introduction of an ad-hoc model. The method is based on matching the mean number density of quasars in haloes with virial mass larger than $M$ to the mean number of quasars more luminous than $L$ for every possible mass and luminosity, i.e.
\begin{equation}
\int_{M}^{\infty} n(M',z)dM' = \int_{L}^{\infty} \phi(L',z) dL'.
\end{equation}
The result is a monotonic relation between halo mass and luminosity, effectively assigning the most luminous quasar to the most massive halo. This is an important assumption, but it is justified by the observation of a direct relation between the quasar luminosity and halo mass \citep{Lidz2006,Chatterjee2012,Shen2009,Shankar2010,ConroyWhite2013} and between the virial halo mass and black hole mass \citep{DiMatteo2003,DiMatteo2005,DiMatteo2008,Shankar2010}.

We use the \citet[][R06 hereafter]{Richards2006} analytical quasar luminosity function to apply the abundance matching. The formula was fitted using the DR5 data, but \citet*{ShenKelly2012} found that this formula is still consistent with DR7 data. Because the formula cannot differentiate between luminosity evolution or number density evolution with redshift, we take advantage of this degeneracy and we assume pure luminosity evolution of the quasar population for the redshift range $1.2 \leq z \leq 1.6$. The R06 luminosity function is
\begin{equation}
\phi(\mathcal{M}_i,z)= \phi^{*} 10^{A\mu}
\end{equation}
where 
\begin{equation}
\mu= \mathcal{M}_i - (\mathcal{M}^{*} +B_1 \xi + B_2 \xi^2 + B_3 \xi^3),
\end{equation}
and
\begin{equation}
\xi \equiv \log_{10} \left( \frac{1+z}{1+z_{\mathrm{ref}}} \right).
\end{equation}
In these equations $\mathcal{M}_i$ is the absolute magnitude in the i-band, and $\phi^{*}$, $A$, $B_1$, $B_2$, $B_3$, $z_{\mathrm{ref}}$ and $\mathcal{M}^{*}$ are free parameters. $z_{\mathrm{ref}}$ is set to $2.45$ and $\mathcal{M}^{*}=-26$. The best fit parameters of the R06 luminosity function for $z \leq 2.4$ are $A=0.84$, $B_1=1.43$, $B_2=36.63$, $B_3=34.39$ and $\log_{10} \phi^{*} = -5.7$.

We obtain $n(<\mathcal{M}_i)$ by integration of $\phi(\mathcal{M}_i,z)$ between $\mathcal{M}_{\rm{b}}$, the absolute magnitude of the brightest quasar detectable in the sample, and $\mathcal{M}_i$, the absolute magnitude of interest. Defining
\begin{equation}
m^*(z)= (\mathcal{M}^{*} +B_1 \xi + B_2 \xi^2 + B_3 \xi^3)
\end{equation}
This results in 
\begin{equation}
\begin{split}
n(<\mathcal{M}_i)= \int_{\mathcal{M}_{\mathrm{b}}}^{\mathcal{M}_i} \phi(\mathcal{M}_i,z) d\mathcal{M}_i =\\ \frac{1}{A \ln(10)} \phi^{*} \left(10^{A(\mathcal{M}_i-m^*(z))}-10^{A(\mathcal{M}_b-m^*(z))} \right).
\end{split}
\end{equation}
We solve the equation for density and impose \\$n(<\mathcal{M}_i)=n(>M_h)$, resulting in
\begin{equation}
\label{eq:mag}
\mathcal{M}_i= \mu + \frac{1}{A} \log_{10}\left[ \frac{A \ln(10)}{\phi^*} n(>M_h) +10^{A(\mathcal{M}_b-\mu)} \right],
\end{equation}
We estimate the cumulative density of haloes with mass $n(>M_h)$ using the rank in mass of each halo divided by the volume of the sample. Then we assign the absolute magnitude using equation \ref{eq:mag} with $\mathcal{M}_b=-29.5$ (the absolute magnitude limit in the sample). Additionally, recent works in HAM acknowledge the possibility of intrinsic scatter in the relation between mass and luminosity in galaxies \citep{Trujillo-Gomez2011,Hearin2013,Behroozi2013}, and they add a random scatter to the relation between mass and luminosity to reproduce better the observations. We follow this procedure and we apply a normal random number with a scatter of 0.3 mag to improve the agreement to 	the observed magnitude distribution. 

Having assigned the absolute magnitudes to the mock quasars, we compute apparent magnitudes using the standard formula in our fiducial cosmology,
\begin{equation}
m_i=M_i+5.0\log_{10}((1+z) D_c)+25.0+K(z),
\end{equation}
where $D_c$ is the comoving distance, $K(z)$ is the K-correction adopted by \citet{Schneider2007},
\begin{equation}
K(z)=-2.5 (1+ \alpha_{\nu}) \log_{10}(1+z),
\end{equation}
and $\alpha_{\nu}$ is the power law index for SDSS-DR7 quasars, which is found to be $\alpha_{\nu}=-0.5$. The radial selection function is then reproduced by applying the magnitude limit of the SDSS-DR7 sample ($m_i<19.1$mag). The resulting redshift distribution follows very well the original radial selection function (see Figure \ref{fig:rad}). 

\subsubsection{Angular selection function}

\begin{figure}
\centering
\includegraphics[width=0.5\textwidth]{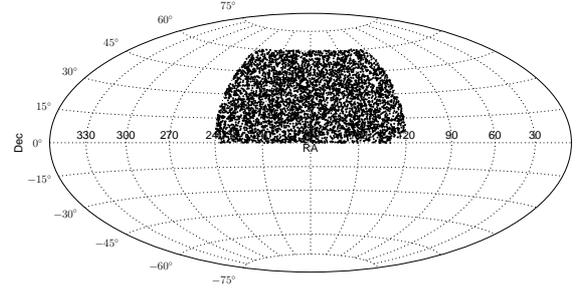} 
\caption[Hammer-Aitoff sky projection of one of the mock quasar samples]{Hammer-Aitoff sky projection of one of the mock quasar samples. The angular distribution in this figure includes the angular selection function of the survey. We display only half of the quasars, selected at random, in this figure. The overall distribution is similar to the observed quasar sample. Compare with Fig. \ref{fig:sampleprojection}. The effect of the angular selection function can be noticed in the lack of quasars in the bottom right corner of the sample distribution.}
\label{fig:mockproj}
\end{figure}

We used the same limits in right ascension and declination for the mock catalogue as in the comparison sample. These take out most of the highly incomplete regions at the boundaries of the observational survey, but the completeness inside this region can vary from point to point, sometimes very strongly, and there are regions with almost no data, mainly because of very bright stars or highly obscured regions. The change in the completeness of the survey with sky position is called the angular selection function. We used the angular selection maps of the SDSS produced by \citet{Blanton2005} using the MANGLE polygon format \citep{Swanson2008,Hamilton2004}. We apply the MANGLE \textit{polyid} command \citep{Swanson2008} to obtain the completeness weights of each mock quasar and then we randomly sample mock quasars with a probability equal to the completeness associated with each mock quasar. The completeness in the angular selection map is defined as the fraction of targets from the photometric survey with measured redshifts. This filtering effectively reproduces the angular selection of the comparison sample, as seen in Fig. \ref{fig:mockproj} compared to its equivalent for the comparison quasar sample (Fig. \ref{fig:sampleprojection}).

\subsubsection{Fibre collisions}

Fibre collisions cause a bias in the quasar sampling in SDSS-QSO DR7 due to the thickness of the fibre cladding, so no two fibres on the same spectroscopic plate can be placed within 55 arcsec of each other \citep{York2000}, corresponding to a spatial separation of $\approx 0.9 h^{-1}$Mpc at the typical quasar redshift of $z = 1.4$. Therefore, the expected number of quasars per halo in the SDSS-QSO sample is one quasar per halo. This effect was considered in the original survey design of the sky sampling, and many regions were observed more than once. However, this repeated sampling was not uniform in the survey area and the procedure was performed for all objects, not just quasars, so in practice the resampling does not correct efficiently the fibre collision bias in the quasar sample. Therefore, there are very few pairs closer than 55 arcsec. In our quasar sample, the number of quasars closer to each other than 55 arcsec is 18, just $0.2$ per cent of the total number of quasar in this sample (10804 quasars). As most of these quasar ``pairs'' are due to line-of-sight projection, we simulate this selection effect by randomly sampling only one quasar in each halo, because the probability of observing quasar pairs in the same halo is negligible. The mean number of quasar in our mocks closer than 55 arcsec after this sampling is $17.53$, with a standard deviation of $5.8$. Thus, our mocks are statistically compatible with observations. These statistics suggest that the observed number of quasar pairs closer than the fibre collision limit in the DR7 catalogue is consistent with line-of-sight projection.

\subsubsection{Redshift distortions}

Observational redshifts are affected by redshift distortions, also called ``Fingers-of-God'' effect, due to the peculiar velocities of quasars. As our mock catalogues include peculiar velocities, we can directly compute the observational redshift. We use the non-relativistic formula
\begin{equation}
z_{\mathrm{obs}}=z_{\mathrm{cos}}+\frac{v_{\mathrm{pec}}}{c} \left(1+z_{\mathrm{cos}}\right),
\end{equation}
where $z_{\mathrm{obs}}$ is the mock observational redshift of the mock quasar, $z_{\mathrm{cos}}$ is the redshift of the quasar computed from its comoving distance, and $v_{\mathrm{pec}}$ is the peculiar velocity assigned to the mock quasar.
 
\subsubsection{Observational errors}

Observational errors can produce many systematic effects in the observational sample that need to be reproduced in the mock catalogues, so that is possible to make sensible comparison. The main variables affected by errors are the redshift and the apparent magnitude. We simulate these errors in our mocks by adding a normally-distributed random variable with a scatter equal to the estimated error in the variables. The estimated observational redshift uncertainty is approximately $0.003$ \citep*{HewettWild2010}, whereas the estimated error in absolute magnitudes for SDSS-DR7 quasars is $0.03$ magnitudes \citep{Schneider2007}.

\subsection{Final construction and testing}

The model just described is inherently probabilistic and therefore it can produce many independent random realizations from the same halo catalogue. This is desirable as it allows us to understand the effect of sampling noise in the LQG catalogue. We run the model 10 times for each of the 11 mock volumes constructed from the HR2 simulation, producing a total of 110 mock quasar catalogues.

\begin{figure}
\centering
\includegraphics[width=0.5\textwidth]{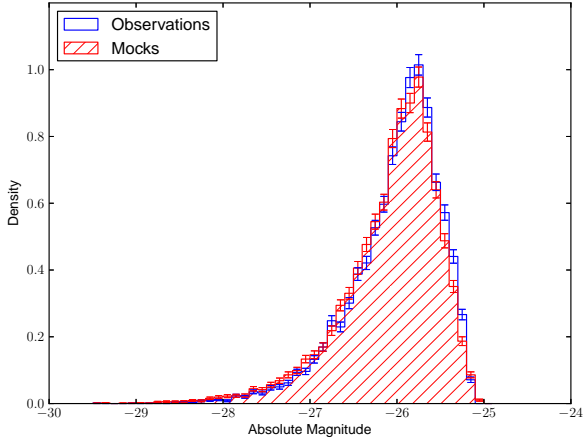}
\caption{Absolute magnitude distributions of SDSS-DR7 quasar (blue) and one of our mock quasar catalogues (red). Both distributions agree well with each other, although there is a small excess of quasars at high luminosity, but this has not affected the following analyses.}
\label{fig:abslum}
\end{figure}

\begin{figure}
\centering
\includegraphics[width=0.5\textwidth]{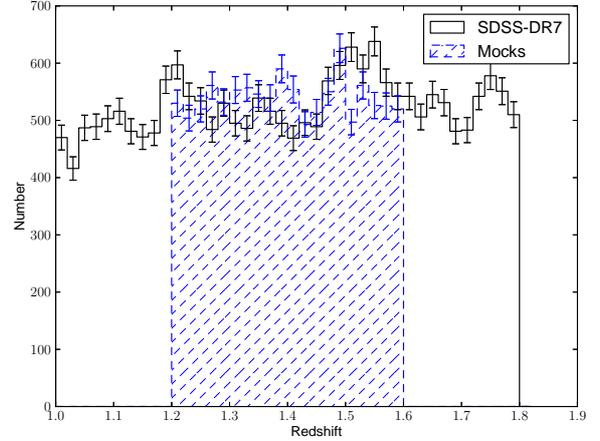}
\caption[Redshift distribution of quasar]{Redshift distribution of quasars in the observational catalogue and in a single mock catalogue. The range of the observational sample is extended to redshifts between $1.0$ to $1.8$ to provide a better visualization of the general trend. The error bars are Poisson errors for each bin count. The mock redshift distribution shows a good agreement with the observational redshift distribution. Other mock catalogues can present some underprediction at high redshift but they are still consistent within the typical error.}
\label{fig:rad}
\end{figure}

We test the accuracy of the reproduction of the observational clustering properties and selection effects in these mock catalogues. The resulting absolute magnitude distribution is shown in Fig. \ref{fig:abslum}, which shows the absolute magnitude distribution for real and mock samples. As expected, by construction the resulting absolute magnitudes is statistically consistent with observations and, therefore, we successfully reproduced the observed radial selection function in our mocks. The redshift distribution is also successfully reproduced, as shown in Fig.\ref{fig:rad}.

\begin{figure}
\centering
\includegraphics[width=0.5\textwidth]{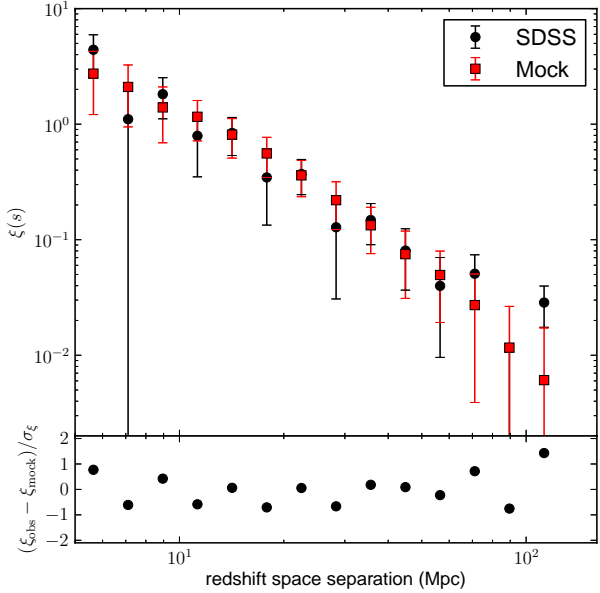}
\caption[The two-point correlation function of SDSS-QSO DR7 and mean correlation function of mock quasar catalogues.]{The two-point correlation function of  SDSS-QSO DR7 and mean correlation function of the mock quasar catalogues. The error bars for SDSS quasars and mocks are the standard deviation from the correlation function computed using all the mocks, therefore including both Poisson or shot noise and cosmic variance. The lower panel shows the standardized residuals (difference between mock and observed correlation functions in each bin divided by the pooled standard error in the mocks $\sigma_{\xi}$). All residuals are within 2 standard errors, indicating that both samples are statistically consistent with the same population hypothesis at 95 per cent confidence level.}
\label{fig:corr}
\end{figure}

To test if the mock quasar catalogues are reproducing the clustering properties of the observational sample, we estimate the two-point correlation function of real and mock quasar catalogues and compare them. We estimate the correlation function in redshift space to include the effect of redshift distortions present in the sample. We use the standard Landy-Szalay estimator \citep*{LandySzalay1993} for the estimation of the two-point correlation function. The estimation of the two-point correlation function requires the construction of a random catalogue to perform the Monte Carlo estimation. The angular coordinates of the random points are obtained using the MANGLE \textit{ransack} command \citep{Swanson2008} using the same angular selection mask that is used in the mock catalogues. This mask indicates the completeness -the fraction of photometric targets successfully include in the redshift catalogue- in each region of the original survey. The radial coordinate is sampled from the normalized radial selection function. The final random catalogues are designed so that the total number of random points is approximately 20 times larger than the total number of quasars in the sample. This higher number of points reduces the sampling error of the estimation of the correlation function. The result is shown in Fig. \ref{fig:corr}. The error bars of both observation and mocks correlation functions are estimated from the standard deviation of the estimates in each bin for the full set of mocks. Using the mock errors for observation error gives a better estimate of the true error as the usual Poisson errors underestimate the true error because they do not take into consideration the cosmic variance. From Fig. \ref{fig:corr}, it is clear that the observed and the mock catalogues are consistent with the observed correlation function of DR7 quasars, within the estimated errors. The mocks are consistent  even at scales below 10 Mpc, when the non-linear evolution begins to be important and the details of the HOD are crucial. The observed deviation in observations at large scales is mostly due to the smaller number of pairs in those bins and the closeness to the scale of the  volume of the sample. Nevertheless, these deviations are within the estimated errors and we find individual mocks with very similar behaviour. The bottom panel of the figure shows the standardized residual, the difference between the observed correlation and the expected correlation in the mocks divided by the pooled standard deviation, i.e. the square root of the sum of the variances of observations and mocks. These show a good agreement between observations and mocks.

We perform a goodness-of-fit test using the statistic $X^2= \sum_{i=0}^N (\xi_{i,\rmn{DR7}}-\xi_{i,\rmn{mock}})^2/\sigma_{\xi,\rmn{mock}}^2$, where $N$ is the number of bins used in the estimation of the correlation function, $\xi$, and $\sigma_{\xi}$ is the standard deviation of $\xi$ in each bin. This statistic is asymptotically distributed as a $\chi^2$ distribution with 13  degrees of freedom. We perform a Pearson $\chi^2$ test to test the null hypothesis that both samples come from the same parent population. The observed statistic is $X^2=11.30$, which is equivalent to a reduced chi-squared statistic $X^2_{\nu}=0.87$ for 13 degrees of freedom ($X^2_{\nu}=X^2/\nu$ where $\nu$ is the number of degrees of freedom of the statistic). The p-value for this statistic is over $0.59$, and therefore it is not possible to reject the hypothesis that both samples come from the same parent population with any reasonable significance level, i.e. the mocks and observations are statistically compatible. Therefore, we conclude that the parameters of the fiducial model are in good agreement with the observed large-scale distribution of quasars in the observational sample.

\section{Mock Large Quasar Groups}
\label{sec:mocklqg}
\begin{figure}
\centering
\includegraphics[width=0.5\textwidth]{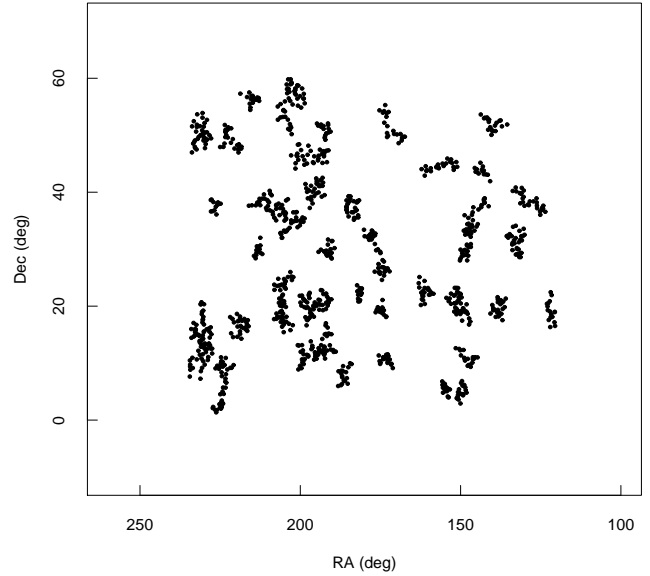}
\caption{RA-Dec plot of the LQG member quasars for a single mock LQG catalogue. The distribution of quasars in each LQG is similar to the distribution of observational LQGs (see Fig. \ref{fig:lqgradec}).}
\label{fig:mocklqgradec}
\end{figure}

We construct the mock LQG catalogues by applying the same procedure described in section \ref{sec:obslqg} to each mock quasar catalogue. We obtained 110 mock LQG catalogues in this way. The mean number of significant LQGs in the mocks is $57.6\pm 0.6$ and the standard deviation in the number of significant LQGs is $6.1 \pm 0.4$. The total number of significant LQGs across all the mocks is $6339$. An example of these mock catalogues is shown in Fig. \ref{fig:mocklqgradec}. We use the summary statistics discussed in section \ref{sec:obslqg} to study the properties of the mock LQGs. The number of members, named quasar number, and characteristic size are particularly important for the study of the largest LQG in the volume as the largest LQG is the maximum in these quantities.

\subsection{Mock large quasar groups properties}

\begin{figure*}
\centering
\includegraphics[width=0.48\textwidth]{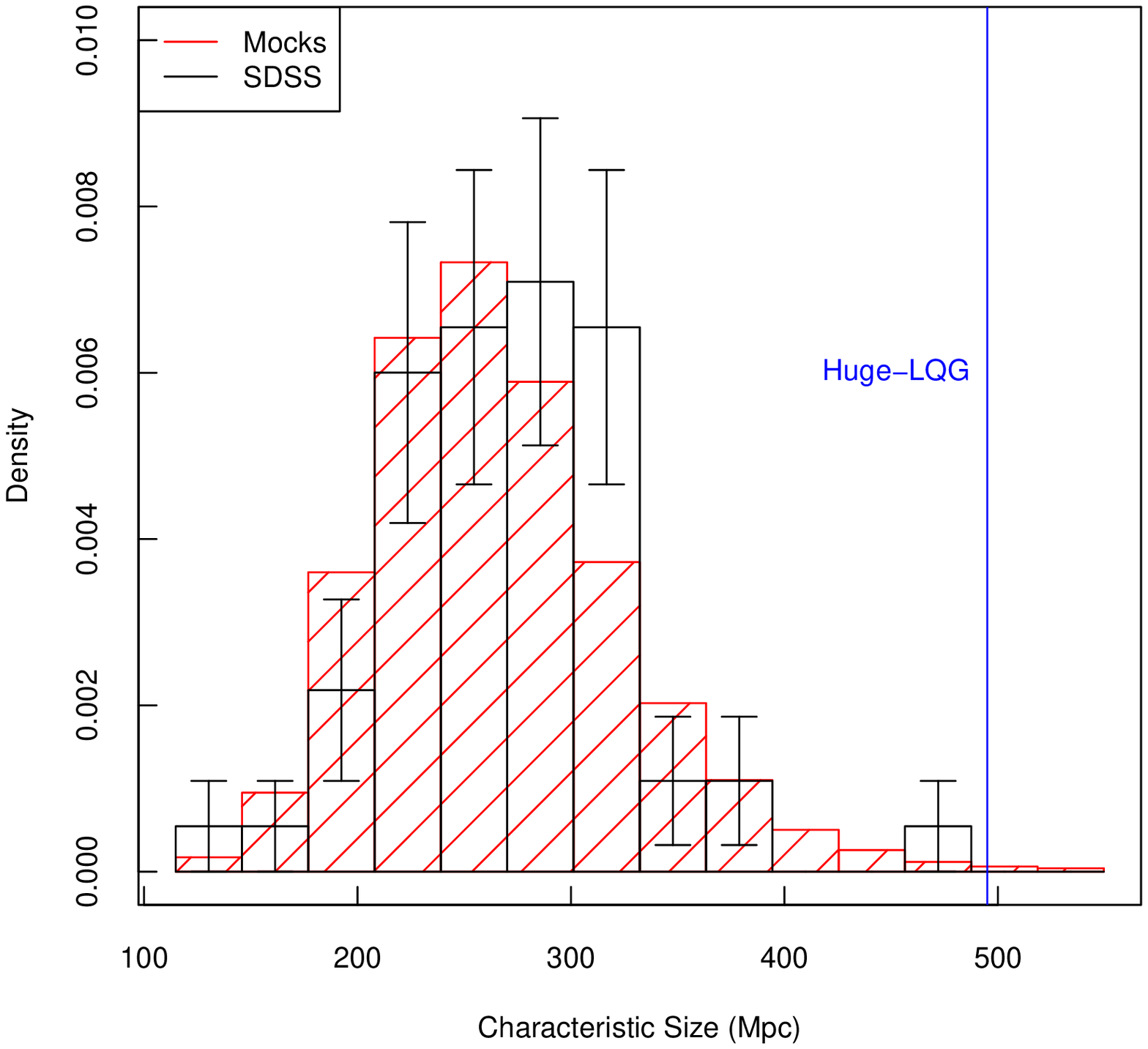}
\includegraphics[width=0.48\textwidth]{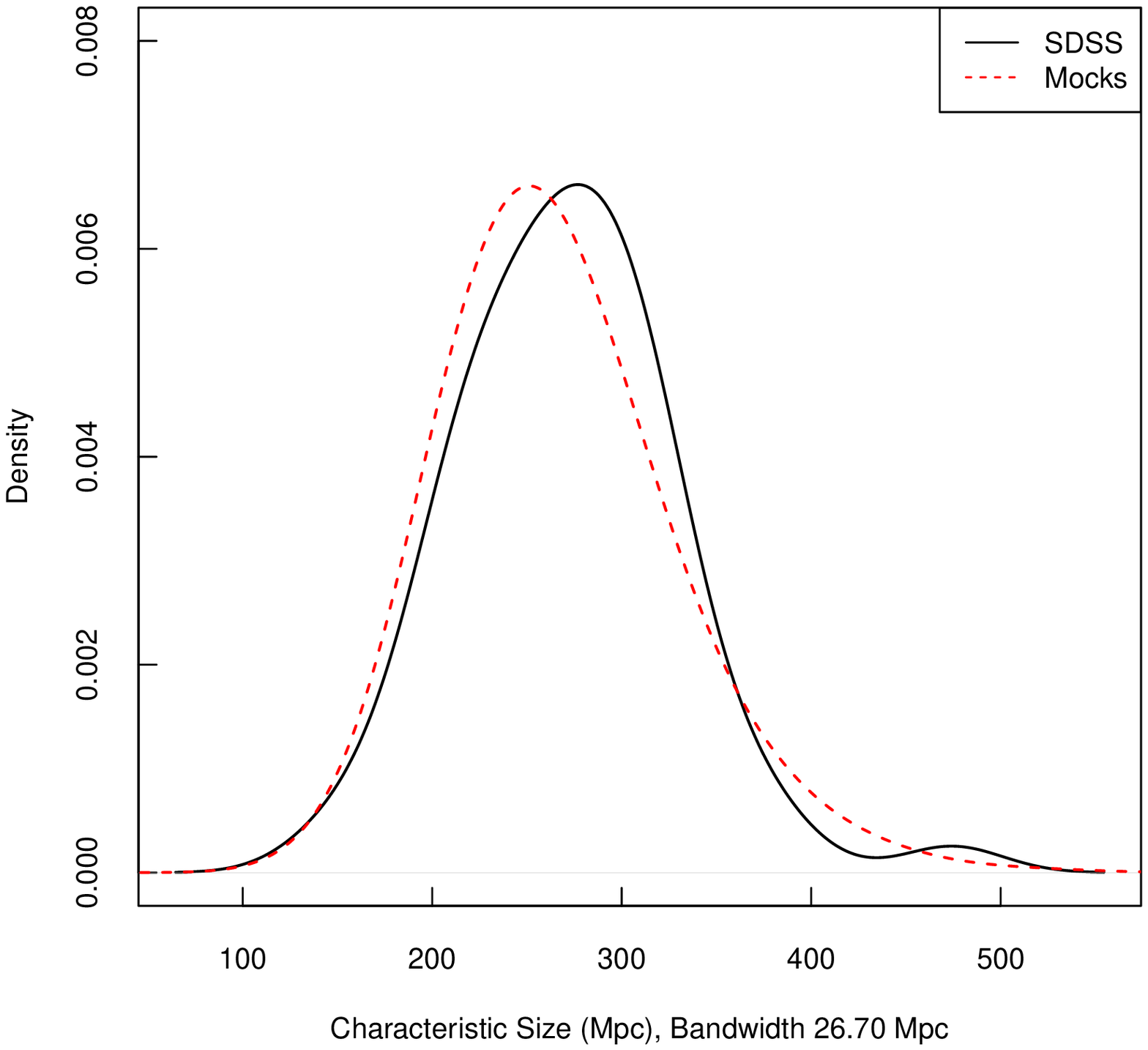}
\caption[Characteristic size for observational sample and stacked mock catalogues.]{Left panel: histograms of the characteristic size distribution for the observational and stacked mock catalogues. Each bin of the observational sample has an error bar showing the Poisson estimate of the error. Both samples are similar, although the observational sample mode seems to be shifted to higher size. Nevertheless, this deviation is compatible with the same parent population hypothesis as tested using a two-sample KS test (p-value $0.52$). The size of the Huge-LQG, as detected in the original catalogue, is shown with a vertical blue line in the left panel. In the current sample the size is somewhat smaller due to the loss of a few members caused by the reduced redshift range, but it is still very close to the original detection. It can be seen that the probability of the Huge-LQG is quite low in the mock catalogues. Right panel: kernel density estimation of the characteristic size distribution for the observational sample and stacked mock catalogues. It provides the same information as the histogram but it does not suffer from the bin location arbitrariness. The bandwidth of $26.70$ Mpc was chosen using Scott's rule-of-thumb \citep{scott2015multivariate}.}
\label{fig:histrealmocksize}
\end{figure*}

The size of a LQG is measured using what we call the characteristic size ($D_\rmn{ch}$), the cube root of the volume estimated with the CHMS method \citep[$V_{\mathrm{CHMS}}$, see][]{Clowes2012}. The characteristic size of mock and SDSS samples is shown in  Fig. \ref{fig:histrealmocksize}. The figure shows the histogram and the kernel density estimation using a Gaussian kernel. The mock sample in this plot is the stacked sample from all 110 mock volumes. Using this procedure it is possible to include the effects of cosmic variance and halo occupation distribution sampling. As the error bars for the stacked sample are very small because of the substantially larger number of LQGs, we do not show the error bars in Fig. \ref{fig:histrealmocksize}. The consistency between both distributions is driven mainly by the error bars of the SDSS LQGs. This is also the case for the other properties. Both distributions are very similar, but there is a slight excess of density at higher sizes in the observations that shift the mean. This shift can be better observed in the kernel density plot. We test if the two samples are consistent with the hypothesis that they come from the same distribution using the two-sample Kolmogorov-Smirnov test (KS test). The p-value for the KS statistic is $0.52$ and, therefore, the samples are consistent with the null hypothesis that they come from the same distribution. However, the KS test is known to be insensitive to discrepancies in the tails of the distribution. In this case alternative tests, like the Cramer or the Anderson-Darling tests, are more powerful. Using the Anderson-Darling 2-sample test \citep{ScholzStephens1987} we also obtain a p-value of $0.52$, indicating that tail discrepancies are not affecting the KS test result.

The size of the Huge-LQG in the original catalogue is marked by a blue vertical line in Fig. \ref{fig:histrealmocksize}. It is clear from the plot that this group is also rare in the mock LQG catalogues. Using the empirical distribution, it is estimated that the probability of finding a group as large or larger than the Huge-LQG is $P(D>495)= 0.003$.

However, this probability cannot be used as a sort of p-value, because it does not take into consideration the size of the sample volume. The volume of the sample defines the size of the LQG sample and if the volume is large enough even unlikely LQG sizes can be sampled with probability one. A better way to assess if the Huge-LQG is an outlier is by using the probability distribution of the largest structure in the sample, i.e. the maximum characteristic size in the volume of the comparison LQG sample. The study of the probability maximum and minimum of a sample is known as extreme value statistics and it is widely used to assess the probability distribution of extreme measurements. We summarise the main aspects of  the extreme value statistics theory in section \ref{sec:eva} and we apply it to characteristic size and quasar number.

\begin{figure*}
\centering
\includegraphics[width=0.48\textwidth]{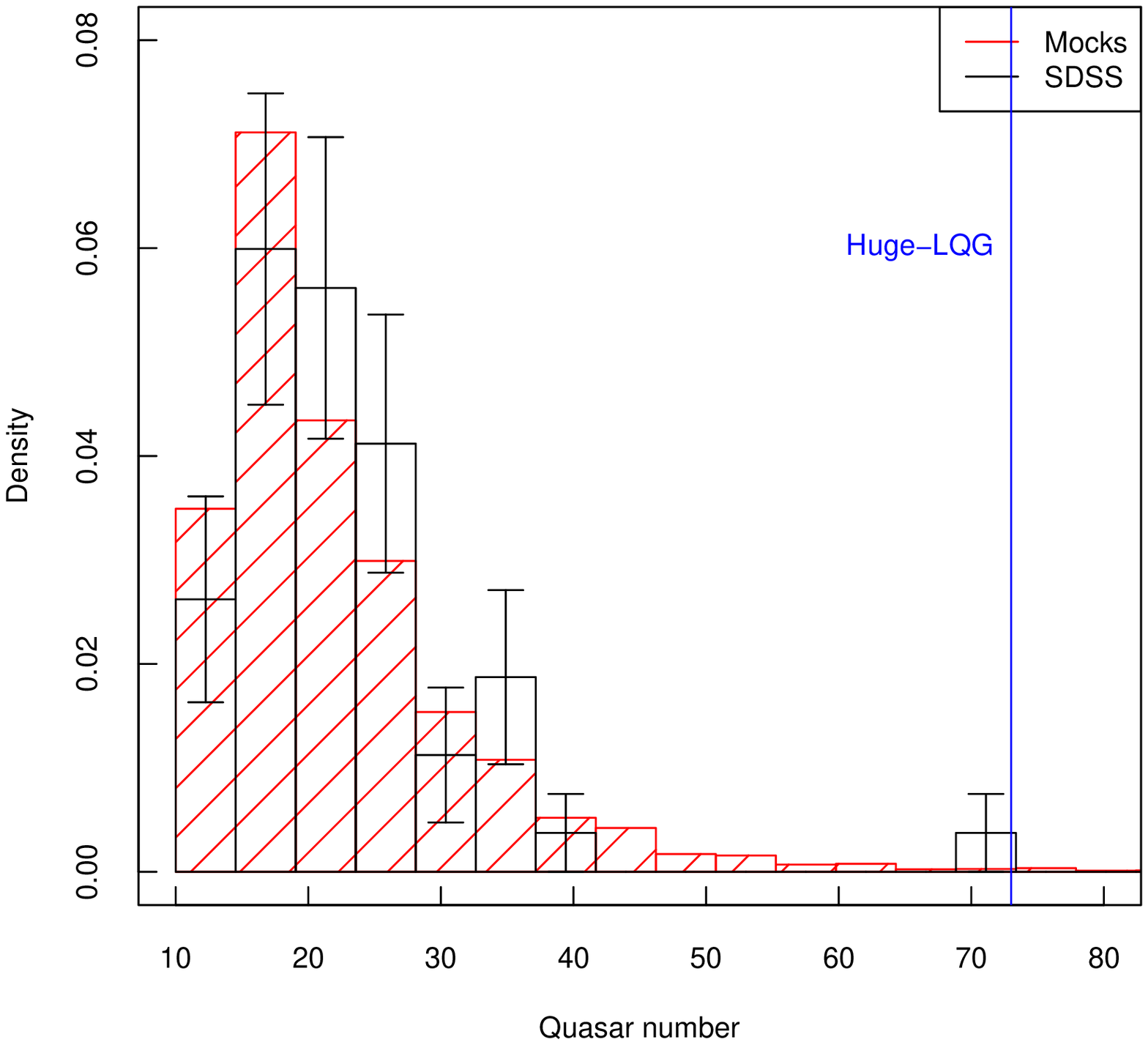}
\includegraphics[width=0.48\textwidth]{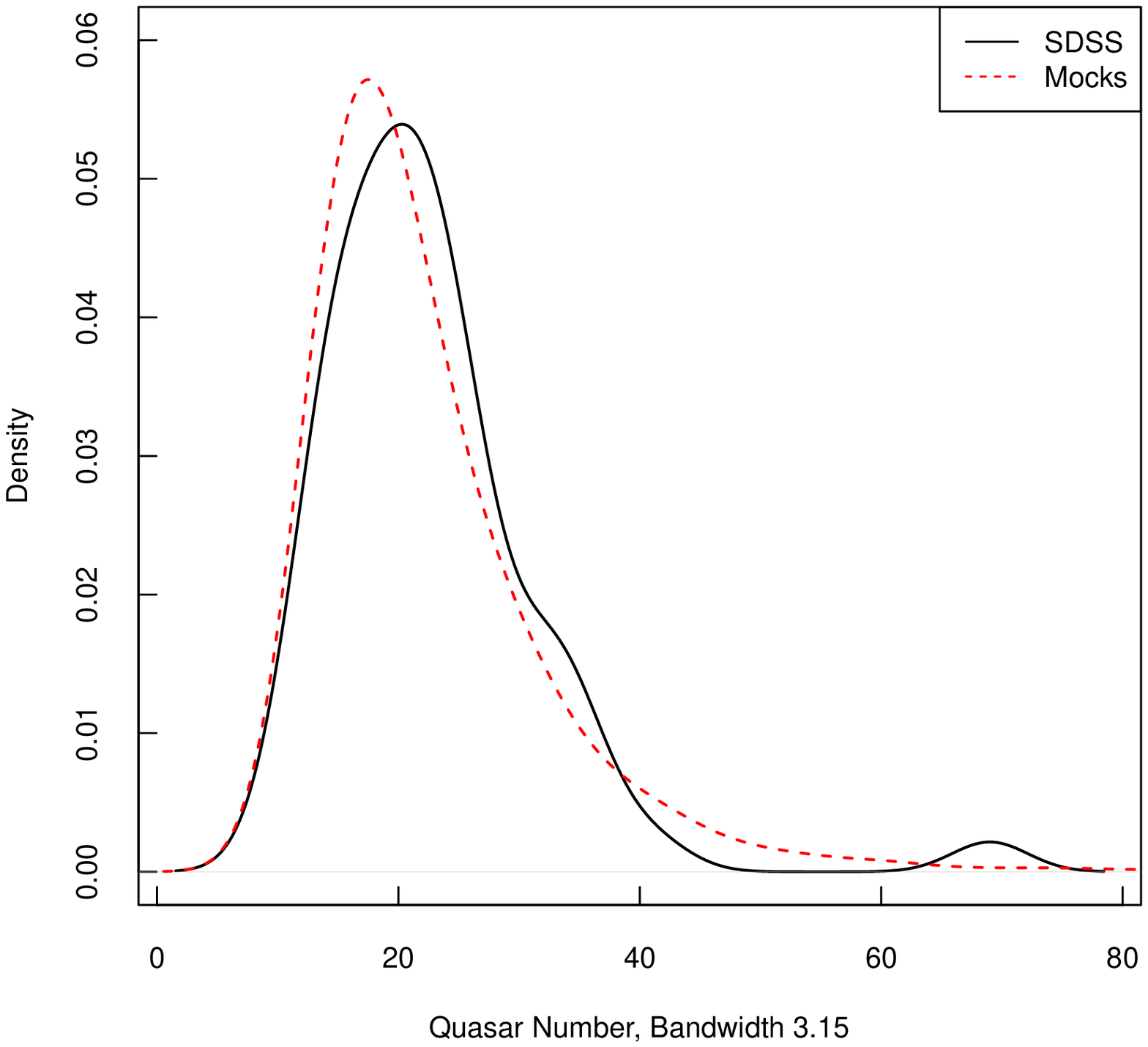}
\caption{Left panel: histograms of quasar number for the observational sample and stacked mock catalogues. Each bin of the observational sample has an error bar showing the Poisson estimate of the error. Both samples have similar distributions. The observational distribution shows more discrete noise in each bin, but this is a consequence of the small number of LQG in each bin. The samples are consistent with the hypothesis of the same parent population tested using the two-sample KS test (p-value $0.73$). The quasar number of the Huge-LQG in the original sample is shown with a vertical line. Again, it is a rare observation according to the stacked mock LQG sample. Right panel: kernel density estimation of quasar number distribution for the observational sample and stacked mock catalogues. We used a Gaussian kernel with a bandwidth of 3.15, chosen using Scott's rule-of-thumb \citep{scott2015multivariate}. The distributions of each sample are very close, except for a drop in the probability density at quasar number higher than 45, which is likely due to the low probability of LQG at the tail of the distribution.}
\label{fig:histrealmocknum}
\end{figure*}

Quasar number is another property in which the Huge-LQG is extreme. This is not a surprise, as both characteristic size and quasar number are related together through the LQG mean quasar density (a larger LQG must have a large number in order to have a similar mean quasar number density). Fig. \ref{fig:histrealmocknum} shows the quasar number distribution of the SDSS and mock LQGs. The quasar number of the Huge-LQG is also shown. The distribution shows an important peak at $\approx 20$ and an exponential decline at large quasar number. The observational sample presents some deviation that is most likely caused by the small number of LQGs in the tail. We apply the two-sample KS test between observation and mocks and it results in a p-value of $0.73$, and therefore we cannot reject the hypothesis that both samples come from the same parent distribution. Using the Anderson-Darling 2-sample test \citep{ScholzStephens1987}, which is more sensitive to differences in the tail of the distributions, we obtain a p-value of $0.55$, again agreeing with the KS test.

As in the case of size, the probability of observing a LQG with more quasars that Huge-LQG is $P(N_q>73)= 0.003$. Therefore, the number of quasar member does not give new information in addition to the characteristic size.

\begin{figure*}
\centering
\includegraphics[width=0.48\textwidth]{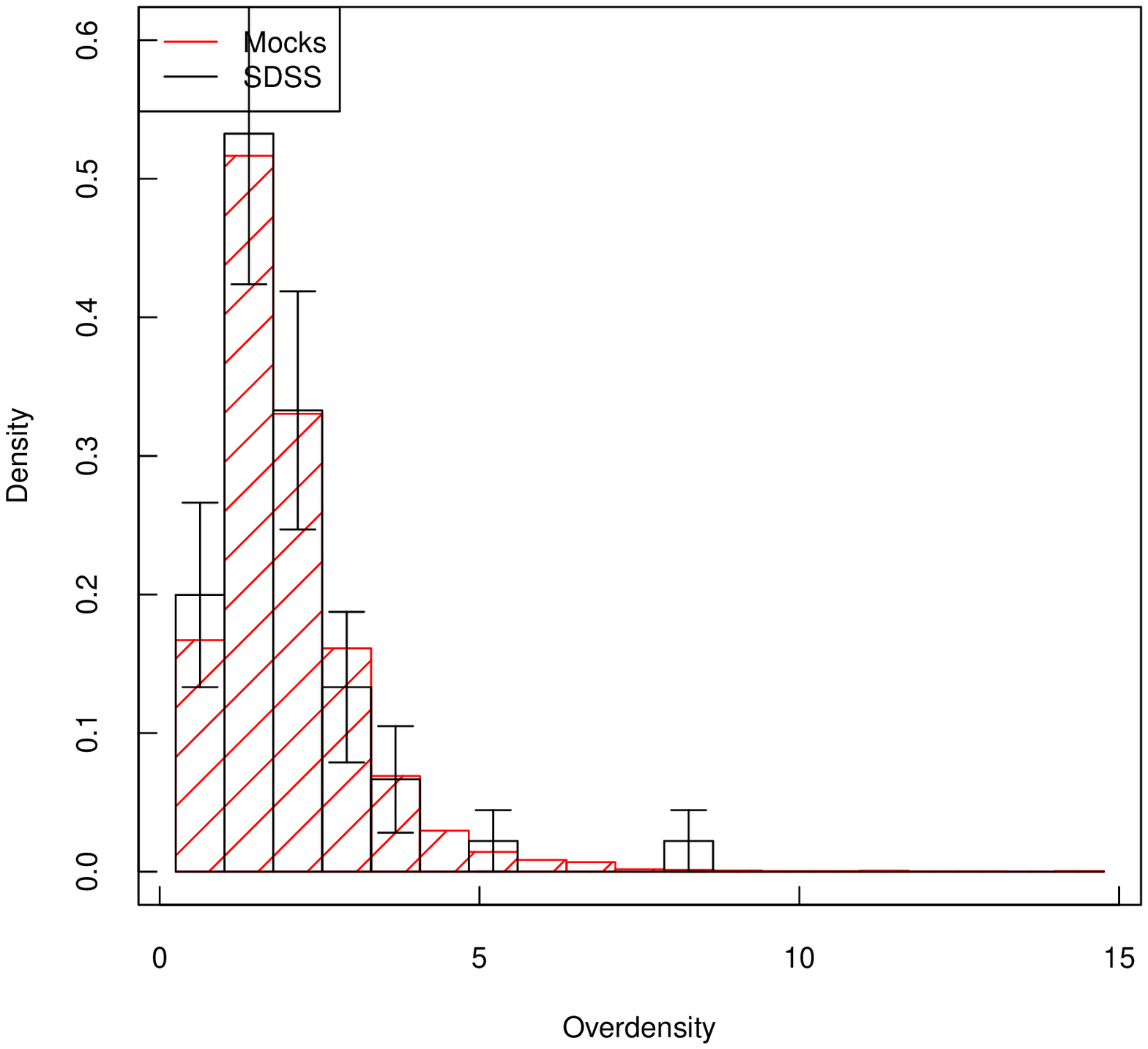}
\includegraphics[width=0.48\textwidth]{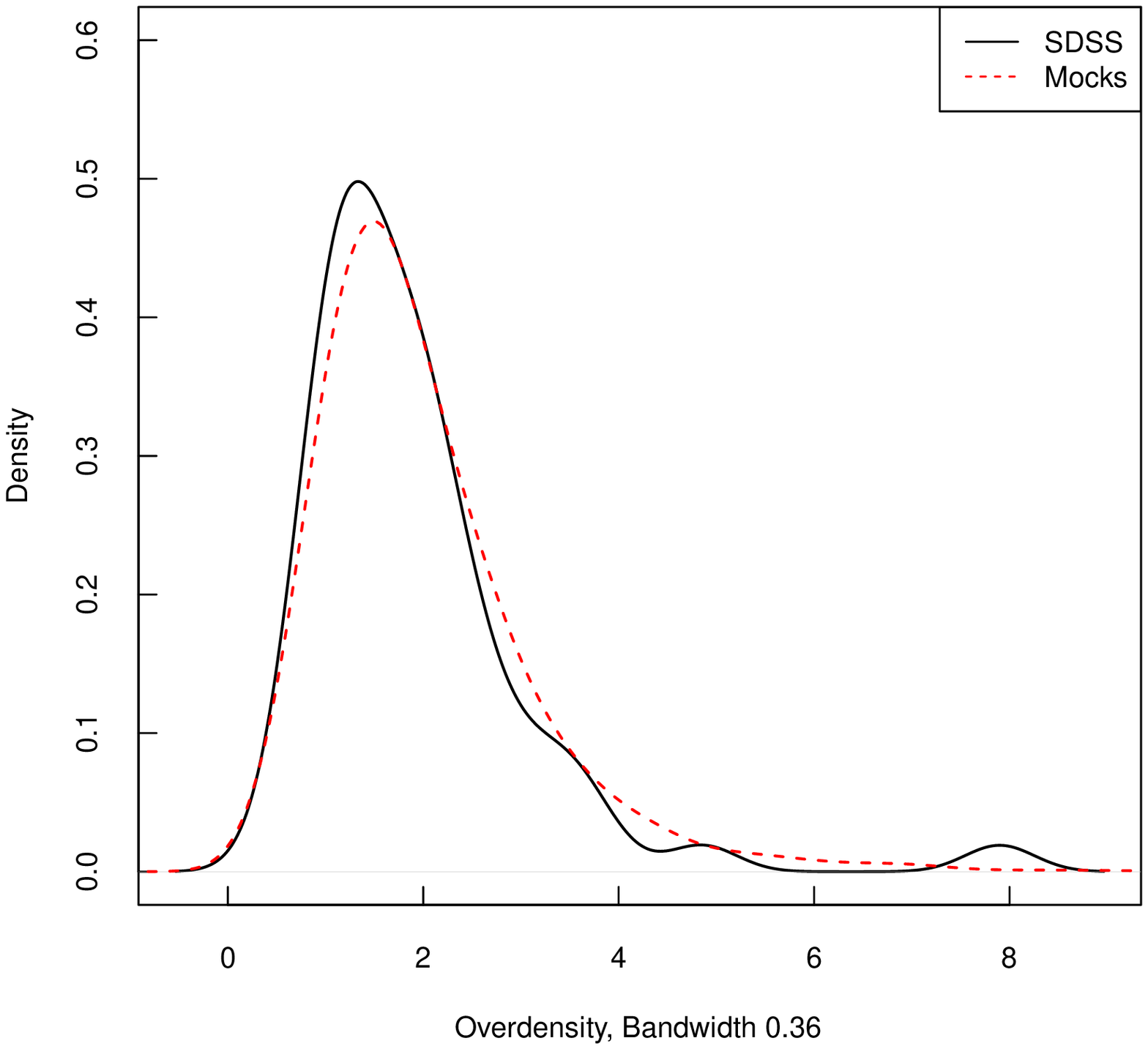}
\caption{Left panel: histogram of overdensity for the observational sample and stacked mock catalogues. The error bars show the Poisson estimate of the error in each bin for the observational sample. Both distributions are very similar and they are compatible with the same parent population hypothesis (under a KS test with p-value $0.62$). The Huge-LQG as a whole is not especially overdense ($\delta_q = 0.4$). Its subsets are more overdense ($\delta_q=1.2$ and $1.54$), but it is still not as overdense as the densest LQG in the sample ($\delta_q = 7.9$), which are usually compact structures with low membership and size. At the right is the kernel density estimation of the overdensity distribution for the observational sample and stacked mock catalogues. We used a Gaussian kernel with bandwidth of 0.36 selected using Scott's rule-of-thumb \citep{scott2015multivariate}. Observational and mock samples show good agreement.}
\label{fig:realmockover}
\end{figure*}

Mean overdensity is an important property of the LQG as it is independent of quasar number and size, which are mainly determined by the approximate threshold density set by the linking length used. The distribution of mean overdensity of the mock LQGs is compatible with SDSS-LQG sample, as can be seen in Fig. \ref{fig:realmockover}. A KS test results in a p-value of $0.62$, therefore both distributions are compatible. It is worth mentioning that in the largest overdensity in the SDSS-LQG catalogue is $7.9$, which is approximately 20 times denser than the Huge-LQG ($\delta_q = 0.4$), therefore we are not performing any extreme value analysis in this quantity. The largest overdensities in the mock LQGs are similar to the one in SDSS-LQG with a mean of $6.9 \pm 0.2$ and standard deviation of $2.0 \pm 0.1$. The largest mean overdensity in the mocks is $14.8$.

\section{Extreme value analysis}
\label{sec:eva}

Extreme value statistics is the study of the distribution of rare events that are produced by the tails of a distribution \citep{coles2001extstat, castillo2004extreme,beirlant2006statistics}. There are two main approaches available. One approach is the study of the maxima and minima of a sample or data block (a period, for example) as the statistic under repeated sampling. This approach is called the block maxima (or minima) method (BM). The analysis of block minima is easily performed by using the inverse of the data variable. Another approach is based on the estimation of the asymptotic probability distribution of the tail of the distribution, i.e. the probability of observing events larger (or smaller) than a certain threshold, which is called the Peak over Threshold (POT) method. The BM method is easy to implement but tends to disregard a large amount of collected data. .The POT method allows more data to be included in the analysis, which improves the accuracy of the estimation of the asymptotic distribution, but it introduces an ad-hoc parameter, the threshold limit, which needs to be selected a priori and thus introduces biases in the estimation of the tail distribution. In comparison, the BM method results in a less accurate but less biased estimation of the probability distribution. The probability distribution of the largest LQG is the distribution of the maxima (size, number) in a data block (sample volume) and therefore we will restrict ourselves to the BM approach.

In extreme value statistics, parametric estimation is highly relevant as non-parametric estimation is inherently biased because the empirical distribution has a zero probability of observing events larger than the largest maximum in a series of samples. For this reason it is desirable to have a parametric distribution that is able to describe the distribution of the maxima or the tail of the distribution,  which gives more accurate inferences about the expected excess probabilities. Fortunately, in the cases where the distribution is unknown, there are analogues of the Central Limit Theorem for extreme values that provide an asymptotic form of the distribution for large samples. If the underlying distribution of the maximum respects some general regularity conditions (the distribution is continuous and twice differentiable and its tail behaves asymptotically as a power law or exponential) then the maximum tends to an asymptotic distribution for large samples, the Generalized Extreme Value distribution (GEV) \citep*{gumbel1958statistics} (In the POT the conditional distribution of events over the threshold tends to the Generalized Pareto Distribution. Additionally, POT can be modelled using the theory of random point processes.)

The GEV distribution is given by
\begin{equation}\label{eq:gev}
F(x_{max} < z)=G(z)=\exp \left[ - \left\lbrace 1 + \gamma \left( \frac{z - \mu}{\sigma} \right) \right\rbrace^{-1/ \gamma}_{+} \right],
\end{equation}
where $\lbrace x \rbrace_+ = \max(x,0)$, $\sigma > 0$ is the scale parameter, and $-\infty < \mu,\gamma < \infty $ are the location parameter and extreme value index. The GEV is a generalization of the three families of distributions for extreme values depending on the sign of the shape parameter, the extreme value index $\gamma$. The heavy-tailed Fr\'{e}chet distribution results from $\gamma > 0$ and the upper-bounded Weibull distribution from $\gamma < 0$. The special case of $\gamma = 0$ is treated by taking the limit as $\gamma \longrightarrow 0$ resulting in the Gumbel distribution \citep{coles2001extstat},
\begin{equation}
G_0(z)=\exp \left[ - \exp \left\lbrace -\left( \frac{z - \mu}{\sigma} \right) \right\rbrace \right],
\end{equation}
where $-\infty < z < \infty $. The Gumbel distribution is especially important as it is the domain of convergence of many standard distributions, such as the normal and lognormal distributions. The conditions for the existence of the asymptotic limit are very general and most standard continuous distributions, i.e. normal, lognormal, cauchy, etc., converge to the GEV.

In practice, however, we do not know if the underlying distribution respects all of the assumptions of the theorems. Also, the convergence with sample size to the asymptotic distribution can be too slow and the maximum distribution can differ from the GEV. Therefore, the GEV is generally used as a fitting model and its adequacy is decided using some form of model selection and goodness-of-fit test.

Useful concepts in the interpretation of extremes or rare events are the return periods or recurrence intervals, and the return levels. A return period or recurrence interval (area or volume) is the expected number of data blocks observed before observing a maximum that exceeds a threshold $z_p$, called the return level, with exceedance probability $p$. Because, the probability of observing a maximum that exceeds the threshold $z_p$ (success) after $n-1$ failures follows a geometric distribution, the return period is $1/p$. Therefore, the recurrence interval is the interval such that the considered event is expected to be exceeded at least once. For example, a exceedance probability of $p = 0.01$ corresponds to a return period or recurrence interval of $100$ in the block units. The return level $z_p$ is computed from the quantile corresponding to $F(z_p)=1-p$, where $F$ is the cumulative distribution of the random variable. The formula for the quantiles $z_p$ of the GEV $G$ (equation \ref{eq:gev}) is given by
\begin{equation}
z_p =  
\begin{cases}
\mu + \frac{\sigma}{\gamma}\left[y_p^{\gamma} -1 \right], & \mathrm{for} \quad \gamma \neq 0, \\
\mu + \sigma \ln y_p , &  \mathrm{for} \quad \gamma = 0.
\end{cases}
\end{equation}
where $y_p=-1/\ln(1-p)$, with $p$ the exceedance probability or inverse return period. In practice we can either (i) fix the return level and its exceedance probability and compute the corresponding recurrence interval, or (ii) fix the recurrence interval (return period) and compute the corresponding return level. In this work, as the basic data block is the survey volume, we use the term recurrence volume. Because the size and membership of the Huge-LQG defines a natural return level we compute the recurrence volume as estimated from our mock LQG simulations.

\subsection{Estimation of GEV parameters}

The estimation of the parameters of the GEV distribution can be accomplished by available standard point estimation methods for distributions, e.g. Maximum Likelihood Estimation (MLE), generalized method of moments, Bayesian estimation, etc. Additionally, there are some special estimation methods based on the quantiles of the distribution (see referenced books for more detail). In this work we use MLE for the estimation of the parameters of the GEV distribution. This approach is equivalent to a Bayesian Maximum a posteriori estimate (MAP) with a flat prior and the Laplace or Normal approximation for the estimation of the posterior distribution. We find the MLE using the Levenberg-Marquardt algorithm for minimization of the negative log-likelihood. Let $z_1,\dots ,z_m$ be the maxima of $m$ independent samples with sample size $n$, then the log-likelihood for the GEV distribution is
\begin{equation}
\begin{split}
l(\mu,\sigma,\gamma | z_1,\dots ,z_m ) = -\sum_{i=1}^{m}\left\lbrace 1 + \gamma \left(\frac{z_i-\mu}{\sigma} \right) \right\rbrace_{+}^{-1/\gamma} \\ - (1+ 1/\gamma) \sum_{i=1}^{m} \ln \left\lbrace 1 + \gamma \left(\frac{z_i-\mu}{\sigma} \right) \right\rbrace_{+} -m \ln \sigma,
\end{split}
\end{equation}
and the log-likelihood of the Gumbel distribution is 
\begin{equation}
\begin{split}
l(\mu,\sigma,\gamma | z_1,\dots ,z_m ) =  -\sum_{i=1}^{m} \exp \left\lbrace - \left(\frac{z_i-\mu}{\sigma} \right) \right\rbrace \\ - \sum_{i=1}^{m} \left(\frac{z_i-\mu}{\sigma} \right) -m \ln \sigma .
\end{split}
\end{equation}

It is recommended to perform some kind of model selection test for the specific case of the Gumbel distribution as the estimation of the parameters of the GEV will most likely have non-zero extreme value index $\gamma$, given that the probability of a single point in the sampling distribution of the estimator of a continuous parameter is zero. We perform a model selection test for a Gumbel distribution, as this has a reduced parameter space, which increases the accuracy of the parameter estimation. We used the likelihood-ratio test, a standard frequentist hypothesis test and model selection based in the Akaike Information Criterion defined as $\mathrm{AIC} = -2 \ln L_\rmn{max} + 2k$, where $\ln L_\rmn{max}$ is the log-likelihood at the MLE and $k$ is the number of parameters, and the Bayesian Information Criterion defined similarly as $\mathrm{BIC} = -2 \ln L_\rmn{max} + k \ln N$, where $N$ the sample size. Using this form for the information criteria one must choose the model with the smaller information criterion.

\subsection{Extreme value analysis of the Huge-LQG}

We construct the block maxima from the mock LQG catalogues by finding the LQG with the maximum property in each mock LQG catalogue. We are using each mock catalogue regardless of which partition of the original cosmological simulation the mock catalogue comes from or which realization of the HOD model is used. In this way the block maxima sample includes the effect of cosmic variance and duty cycle random sampling, which is necessary to avoid any bias. We consider the two extreme properties of LQGs: characteristic size and quasar number.

\begin{table}
\caption{Maximum Likelihood Estimation for GEV and Gumbel distributions for the characteristic size.}
\label{tab:fitsize}
\begin{center}
\begin{tabular}{l|l}
\hline
\multicolumn{2}{c}{Parameter estimates for GEV}\\
\hline
location $\mu$ & $430.3 \pm 4.5$ Mpc\\
scale $\sigma$ & $41.8 \pm 3.3$ Mpc\\
extreme value index $\gamma$ &  $-0.06 \pm 0.07$\\
AIC & $1168.156$ \\
BIC & $1176.257$ \\
\hline
\multicolumn{2}{c}{Parameter estimates for Gumbel}\\
\hline
location $\mu$ & $429.0 \pm 4$ Mpc\\
scale $\sigma$ &  $41 \pm 3$ Mpc\\
AIC & $1166.712$\\
BIC & $1172.113$\\
\hline
\multicolumn{2}{c}{Likelihood ratio test} \\
\hline
Deviance ($-2 \ln(L_0/L_1)$) & $0.5569$\\
Asymptotic p-value & $0.4555$ \\
\hline
\end{tabular}
\end{center}
\end{table}

\begin{figure*}
\centering
\includegraphics[width=0.48\textwidth]{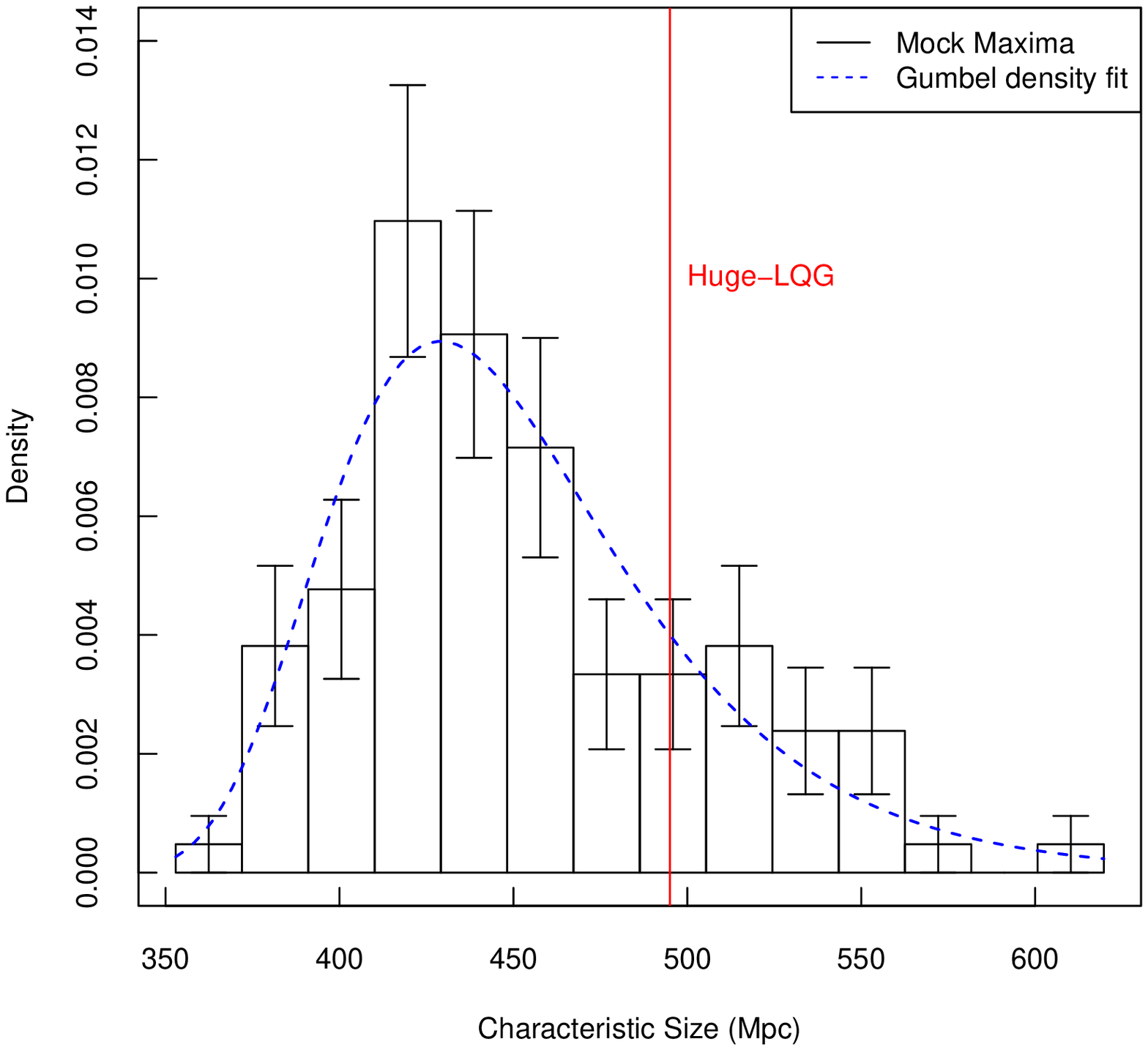}
\includegraphics[width=0.48\textwidth]{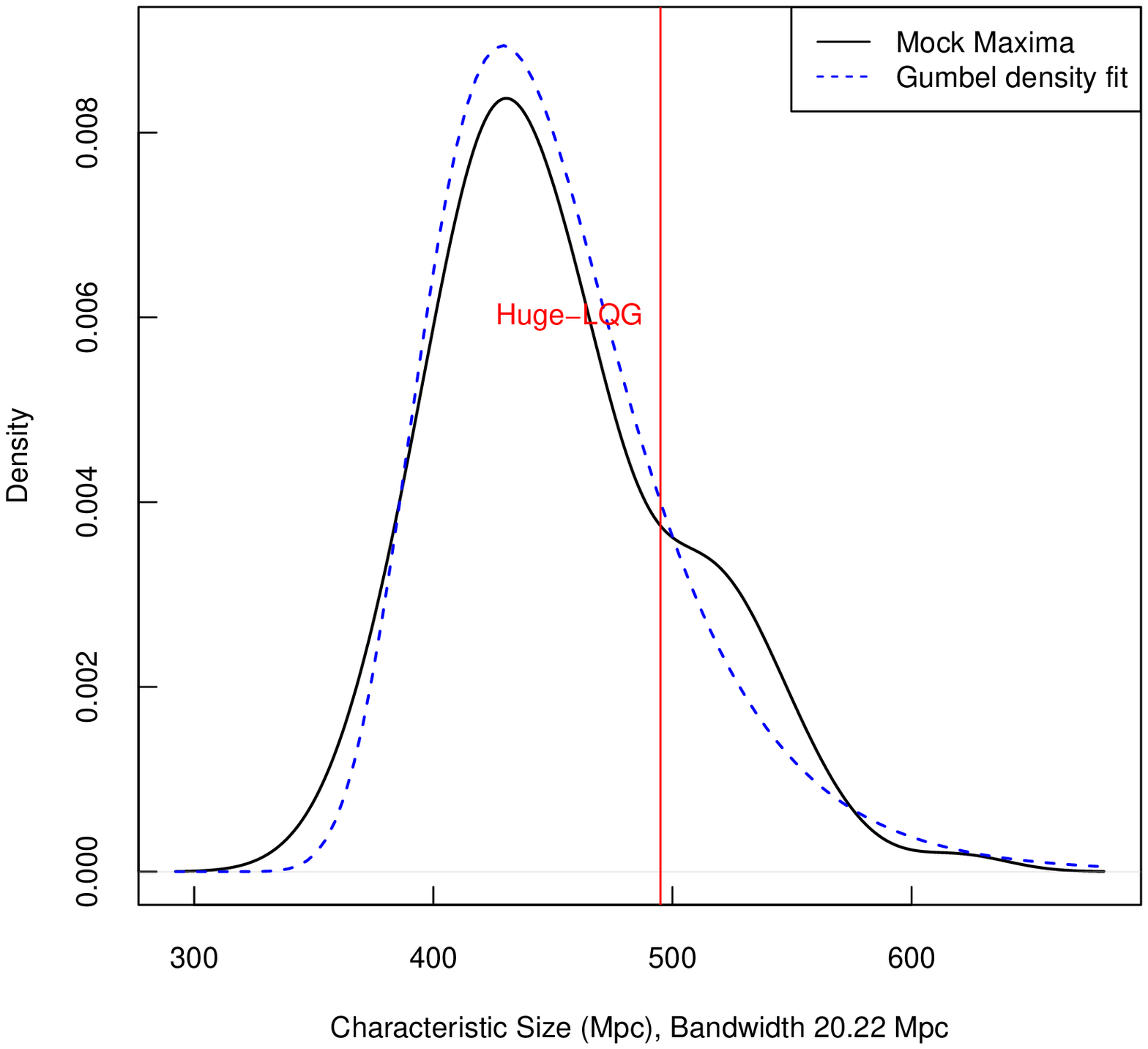}
\caption{Left panel: the histogram of the maximum characteristic size distribution. The error bars in the histogram show the Poisson estimate of the error in each bin. The best fit of the extreme value distribution of the maximum size of the mock sample is shown in the figure with a blue dashed line. The extreme value distribution provides a good fit to the mock maximum distribution. The Huge-LQG observed value for the characteristic size in the original catalogue is shown with a red vertical line. The p-value is very high ($0.19$), and therefore the Huge-LQG size is a common value. Right panel: the kernel density estimation of the maximum characteristic size distribution. We used a Gaussian kernel with a bandwidth of $20.22$ Mpc selected using Scott's rule-of-thumb \citep{scott2015multivariate}. The smoothed distribution of the mock maxima in size shows excellent agreement with the fitted extreme value distribution. There is some oscillation in the large size tail, but these are consistent within the errors.}
\label{fig:extsize}
\end{figure*}

We used the empirical distribution of the size maxima of the mock LQG to provide an initial non-parametric estimation of the size maxima distribution. The probability of observing a LQG with size maxima larger than Huge-LQG ($495$Mpc) is $P(D^\rmn{max}>495$Mpc$) = 0.227$. Therefore, it is possible to say that the Huge-LQG size is not an unlikely maximum in the size-maxima distribution. However, the empirical distribution is a biased estimator of the extreme value distribution (EVD) and the tail might be affected by small sample statistics. A fit to the EVD will produce a better estimation of the p-value of the sample. Fig. \ref{fig:extsize} shows the distribution of maxima size from the mock LQG using a histogram and a density kernel estimate respectively.

We fit the GEV and Gumbel distributions to the size maxima using MLE (see Table \ref{tab:fitsize}). The standard errors are estimated from the inverse of the observed Fisher information. The GEV $1\sigma$ confidence region of $\gamma$ includes zero and therefore we should accept the Gumbel distribution under a Wald test. We performed a likelihood-ratio test with the null hypothesis that the sample comes from a Gumbel distribution and the alternative hypothesis that it comes from a GEV. The test statistic is within the 5 per cent significance level acceptance region of the null hypothesis, thus we should accept the Gumbel distribution as the fitting distribution. The AIC and BIC for the Gumbel are both smaller indicating support for this model. However, the difference is very small in absolute terms. In a Bayesian approach the difference of BIC $\Delta BIC = BIC_1 - BIC_0$ is asymptotically equivalent to -2 times the logarithm of the Bayes factor. The BIC difference in this case is $4.144$, which is considered positive evidence in support of the Gumbel model according to the scale of \citet{KassRaftery1995}.

\begin{figure*}
\label{fig:extnum}
\centering
\includegraphics[width=0.48\textwidth]{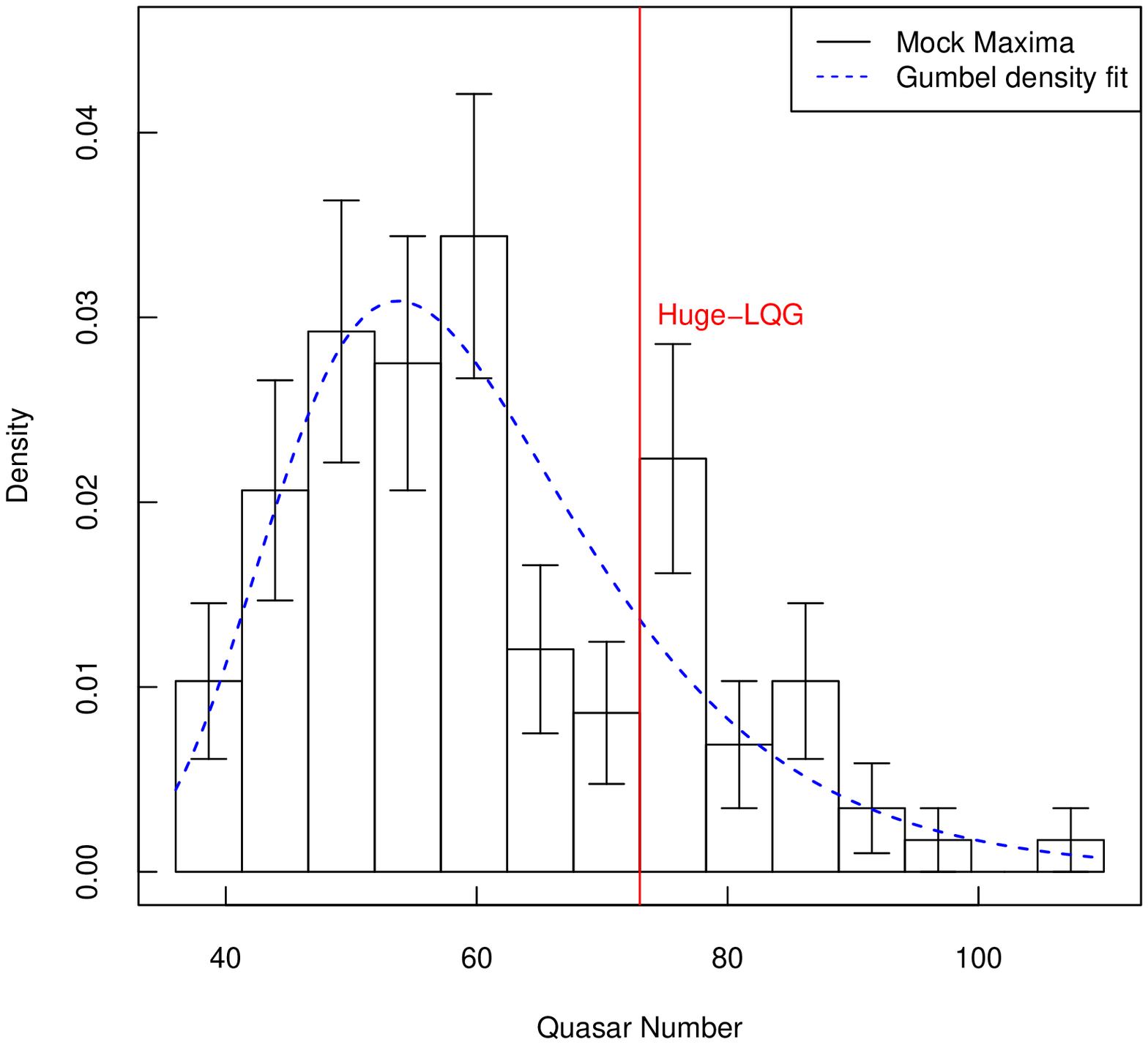}
\includegraphics[width=0.48\textwidth]{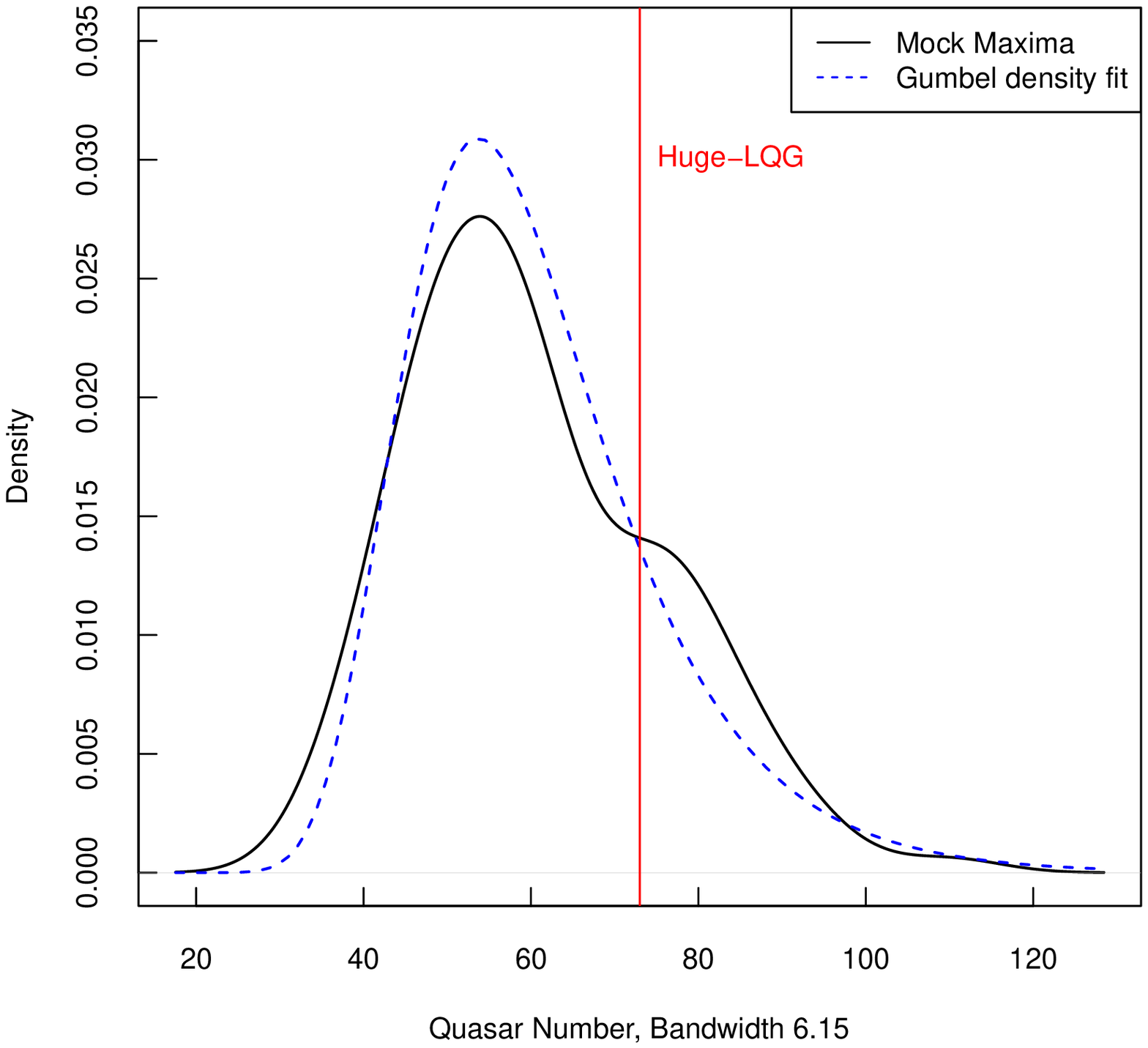}
\caption{Left panel: histogram of the maximum quasar number distribution. The error bars in the histogram show the Poisson estimate of the error in each bin. The best fit of the extreme value distribution to the maximum quasar number in the mock sample is shown in the figure with a blue dashed line. The extreme value distribution provides a good fit to the mock maximum distribution. The Huge-LQG observed quasar number in the original catalogue is shown with a red vertical line. The p-value is very high ($0.18$), and therefore the Huge-LQG quasar number is a common value. Right panel: the kernel density estimation of the maximum quasar number distribution. We used a Gaussian kernel with a bandwidth of 6.15 selected using the Scott's rule-of-thumb \citep{scott2015multivariate}. The extreme value distribution best fit is also shown (blue dashed curve). The Huge-LQG quasar number (red vertical line) is shown for comparison.}
\end{figure*}

It is interesting to compare the fitted distribution to the histogram and the probability density obtained by Kernel Density Estimation (see Fig. \ref{fig:extsize}). Overall, the fitted Gumbel distribution is a good model for mock size maxima. There is, however, a secondary peak in the histogram for LQGs close to the Huge-LQG size, although this is not significant as the error bars show. When observed in the kernel density estimate the secondary peak is also observed. This suggests a possible preference for this kind of LQG as maximum, but it is not possible to confirm this claim using the current data.

The probability of a LQG larger than Huge-LQG, i.e. the exceedance probability, is $P(D^\rmn{max}>D_\rmn{ch}^\rmn{max}=495$Mpc$)=1-G(D_\rmn{ch}^\rmn{max})= 0.19$, similar to the estimation using the empirical distribution. We can estimate the recurrence volume using the inverse of this excedaance probability ($1/p$) that gives as the result $5.3 \pm 1$ volumes, where the error is estimated using Monte Carlo simulation from the best fit distribution (also called parametric bootstrapping). Therefore, the Huge-LQG is a common size maximum in our mock catalogue and it is compatible with the concordance cosmology provided that there is not a similar or larger structure in a survey five times larger.

\begin{table}
\caption{Maximum Likelihood Estimation for GEV and Gumbel distributions for the quasar number.}
\label{tab:fitnum}
\begin{center}
\begin{tabular}{l|l}
\hline
\multicolumn{2}{c}{Parameter estimates for GEV}\\
\hline
location $\mu$ & $53.9 \pm 1.3$\\
scale $\sigma$ & $12.0 \pm 1.0$\\
extreme value index $\gamma$ &  $-0.03 \pm 0.08$\\
AIC & $897.9785$ \\
BIC & $906.0799$ \\
\hline
\multicolumn{2}{c}{Parameter estimates for Gumbel}\\
\hline
location $\mu$ & $53.7 \pm 1.2$ \\
scale $\sigma$ & $11.9 \pm 0.9$ \\
AIC & $896.0828$\\
BIC & $901.4837$\\
\hline
\multicolumn{2}{c}{Likelihood ratio test} \\
\hline
Deviance ($-2 \ln(L_0/L_1)$) & $0.1043$\\
Asymptotic p-value & $0.7467$ \\
\hline
\end{tabular}
\end{center}
\end{table}

The quasar number is another property in which the Huge-LQG is extreme, so despite this property being correlated with the size, it is worth testing if quasar number maxima gives the same information as size. We construct a mock quasar number maxima sample in the same way as for characteristic size and we perform MLE fitting using GEV and Gumbel distributions. Again, the quasar number maxima is compatible with a Gumbel distribution under a likelihood-ratio test and Wald test (confidence region defined by the standard error includes zero) and the AIC and BIC of the Gumbel are smaller than for the GEV. A difference in BIC of 4.6 indicates positive evidence for the Gumbel distribution \citep[][]{KassRaftery1995}. Therefore, we choose the Gumbel distribution as the best model. The resulting MLE for the Gumbel distribution is shown in Fig. \ref{fig:extnum}, and the MLE point estimates are shown in Table \ref{tab:fitnum}.

The probability of observing a LQG with a quasar number larger than Huge-LQG ($N_q=73$) is $P(N_q^{\rmn max}=N_{q,\mathrm{SDSS}}^{\rmn max}=73) = 1-G(N_{q,\rmn{SDSS}}^\rmn{max})=0.18$, which is consistent with the estimation from maximum characteristic size. This implies a recurrence volume of 5.6 volumes. In this sense, the quasar number maxima give the same information about the likelihood of this structure as the characteristic size. A secondary peak at the Huge-LQG quasar number is observed (see Fig. \ref{fig:extnum}), similar to the one observed in the maximum in characteristic size. The Gumbel distribution is still consistent with the mock distribution given the estimated uncertainties.

\section{Conclusions}

The comparison of the mock LQGs with observational SDSS LQGs shows that the distribution of different properties is compatible with the hypothesis of the same population. Therefore, the LQG population in the observational LQG sample is consistent with the expected LQGs in the concordance cosmological model. The Huge-LQG, which was considered a probable outlier of the LQG size and quasar number distribution, is indeed a very rare object in the mocks with approximately 0.3 per cent probability to observe an object larger than this (1 in 330 approximately). However, the extreme value analysis of the mocks shows that the probability of observing an LQG larger than the Huge-LQG (either in size or quasar number) in the sample volume is $\sim 20$ per cent, and therefore the Huge-LQG is not an unlikely maximum. Therefore, we conclude that the Huge-LQG is compatible with the standard cosmology (i.e. as predicted by our quasar model), provided that there is not a similar or larger structure in a survey five times larger containing the current sampled volume.

Our conclusion relies on the adopted model of quasar occupation, which makes some important simplifications on the quasar duty cycle. We find that quasar model is accurate enough for our analysis given the good agreement between the mock and observed LQGs, giving support to the adopted approximations. Nevertheless, our statistics have an important level of uncertainty, mainly caused by Poisson or shot noise caused in turn by the small number densities in the quasar sample. These errors can mask the difference between the theoretical predictions and observations. Therefore, further progress on the analysis of the quasar distribution requires the reduction of these uncertainties, and specifically a larger number of quasars, which requires a new generation of larger and deeper quasars surveys. Also, the formulation of more robust group finders and geometry statistics could be valuable for a more powerful evaluation of the predictions of the cosmological model.

\section*{Acknowledgements}
We gratefully thank Prof. Juhan Kim for providing the Horizon Run 2 N-body simulation dark matter halo catalogues employed in this work. Gabriel Marinello gratefully acknowledges the financial support of the University of Central Lancashire.

This research has used the SDSS DR7QSO catalogue (Schneider et al. 2010). Funding for the SDSS and SDSS-II has been provided by the 
Alfred P. Sloan Foundation, the Participating Institutions, the National Science Foundation, the US Department of Energy, the National Aeronautics and Space Administration, the Japanese Monbukagakusho, the Max Planck Society and the Higher Education Funding Council for England. The SDSS website is http://www.sdss.org/. The SDSS is managed by the Astrophysical Research Consortium for the Participating Institutions. The Participating Institutions are the American Museum of Natural History, Astrophysical Institute Potsdam, University of Basel, University of Cambridge, Case Western Reserve University, University of Chicago, Drexel University, Fermilab, the Institute for Advanced Study, the Japan Participation Group, Johns Hopkins University, the Joint Institute for Nuclear Astrophysics, the Kavli Institute for Particle Astrophysics and Cosmology, the Korean Scientist Group, the Chinese Academy of Sciences (LAMOST), Los Alamos National Laboratory, the Max-Planck-Institute for Astronomy (MPIA), the Max-Planck Institute for Astrophysics (MPA), New Mexico State University, Ohio State University, University of Pittsburgh, University of Portsmouth, Princeton University, the United States Naval Observatory and the University of Washington.

\bibliographystyle{mnras}
\bibliography{biblio}  

\bsp
\label{lastpage}
\end{document}